\documentclass[aps,prb,preprint,superscriptaddress,nobibnotes]{revtex4-1}
\usepackage{graphicx}
\usepackage{epstopdf}
\usepackage{color}
\usepackage{float}
\usepackage[caption = false]{subfig}
\begin{document}

%\preprint{}


\title{Upconverted electroluminescence via Auger scattering of interlayer excitons in van der Waals heterostructures}


\author{J. Binder}

\affiliation{Laboratoire National des Champs Magnetiques Intenses,
CNRS-UGA-UPS-INSA-EMFL, 25 Rue des Martyrs, 38042 Grenoble, France}
\affiliation{Faculty of Physics, University of Warsaw, ul. Pasteura 5, 02-093 Warsaw, Poland}

\author{J. Howarth}
\affiliation{School of Physics and Astronomy, University of Manchester, Oxford Road, Manchester M13 9PL, UK}
\affiliation{National Graphene Institute, University of Manchester, Oxford Road, Manchester, M13 9PL, UK}

\author{F. Withers}
\affiliation{Centre for Graphene Science, College of Engineering, Mathematics and Physical Sciences, University of Exeter, Exeter EX4 4QF, UK}

\author{M. R. Molas}
\affiliation{Laboratoire National des Champs Magnetiques Intenses,
CNRS-UGA-UPS-INSA-EMFL, 25 Rue des Martyrs, 38042 Grenoble, France}
\affiliation{Faculty of Physics, University of Warsaw, ul. Pasteura 5, 02-093 Warsaw, Poland}

\author{T. Taniguchi}
\author{K. Watanabe}
\affiliation{National Institute for Materials Science, 1-1 Namiki, Tsukuba 305-0044, Japan}

\author{C. Faugeras}
\affiliation{Laboratoire National des Champs Magnetiques Intenses,
CNRS-UGA-UPS-INSA-EMFL, 25 Rue des Martyrs, 38042 Grenoble, France}

\author{A. Wysmolek}
\affiliation{Faculty of Physics, University of Warsaw, ul. Pasteura 5, 02-093 Warsaw, Poland}

\author{M. Danovich}
\affiliation{School of Physics and Astronomy, University of Manchester, Oxford Road, Manchester M13 9PL, UK}
\affiliation{National Graphene Institute, University of Manchester, Oxford Road, Manchester, M13 9PL, UK}

\author{V. I. Fal'ko}
\affiliation{School of Physics and Astronomy, University of Manchester, Oxford Road, Manchester M13 9PL, UK}
\affiliation{National Graphene Institute, University of Manchester, Oxford Road, Manchester, M13 9PL, UK}
\affiliation{Henry Royce Institute for Advanced Materials, M13 9PL, Manchester, UK}
\author{A. K. Geim}
\author{K. S. Novoselov}
\affiliation{School of Physics and Astronomy, University of Manchester, Oxford Road, Manchester M13 9PL, UK}
\affiliation{National Graphene Institute, University of Manchester, Oxford Road, Manchester, M13 9PL, UK}

\author{M. Potemski}
\thanks{corresponding author}
\email{marek.potemski@lncmi.cnrs.fr}
\affiliation{Laboratoire National des Champs Magnetiques Intenses,
CNRS-UGA-UPS-INSA-EMFL, 25 Rue des Martyrs, 38042 Grenoble, France}
\affiliation{Faculty of Physics, University of Warsaw, ul. Pasteura 5, 02-093 Warsaw, Poland}

\author{A. Kozikov}
\thanks{corresponding author}
\email{aleksey.kozikov@manchester.ac.uk}
\affiliation{School of Physics and Astronomy, University of Manchester, Oxford Road, Manchester M13 9PL, UK}
\affiliation{National Graphene Institute, University of Manchester, Oxford Road, Manchester, M13 9PL, UK}


%\date{\today}

\begin{abstract}
The intriguing physics of carrier-carrier interactions, which likewise affect the operation of light emitting devices, stimulate the research on semiconductor structures at high densities of excited carriers, a limit reachable at large pumping rates or in systems with long-lived electron-hole pairs. By electrically injecting carriers into WSe$_2$/MoS$_2$ type-II heterostructures which are indirect in real and k-space, we establish a large population of typical optically silent interlayer excitons. Here, we reveal their emission spectra and show that the emission energy is tunable by an applied electric field. When the population is further increased by suppressing the radiative recombination rate with the introduction of an hBN spacer between WSe$_2$ and MoS$_2$, Auger-type and exciton-exciton annihilation processes become important. These processes are traced by the observation of an up-converted emission demonstrating that excitons gaining energy in non-radiative Auger processes can be recovered and recombine radiatively.\end{abstract}

%\pacs{}
%\keywords{}

\maketitle


\section*{Introduction}

Type-II heterostructures made of semiconducting monolayer transition metal dichalcogenides (TMDs) are excellent systems to study indirect interlayer excitons and their properties, in particular at high charge carrier densities. In the regime of large carrier injection many phenomena and processes become important that are negligible at low injection rates \cite{Pandey2007,Qian2010,He2016}. One of those processes is Auger recombination, a non-radiative decay of non-equilibrium carriers, inherent to any semiconducting material. Since for the Auger process to proceed at least three charge carriers are required, the process gains importance with increasing carrier densities. Notably, Auger-type processes become also a factor in limiting the performance of light-emitting diodes (LEDs) at high output power\cite{Iveland2013}. The importance of Auger-type exciton-exciton scattering processes was evidenced for direct intralayer excitons in semiconducting TMD monolayers by optical pump probe and time-resolved photoluminescence experiments \cite{Sun2014,Yuan2015,Poellmann2015,Yu2016,Aivazian2017,Kulig2018}. Here, we study indirect interlayer excitons in electroluminescent devices with a particular focus on the high carrier density regime.

\begin{figure}
	\centering
		\includegraphics[width=1\textwidth]{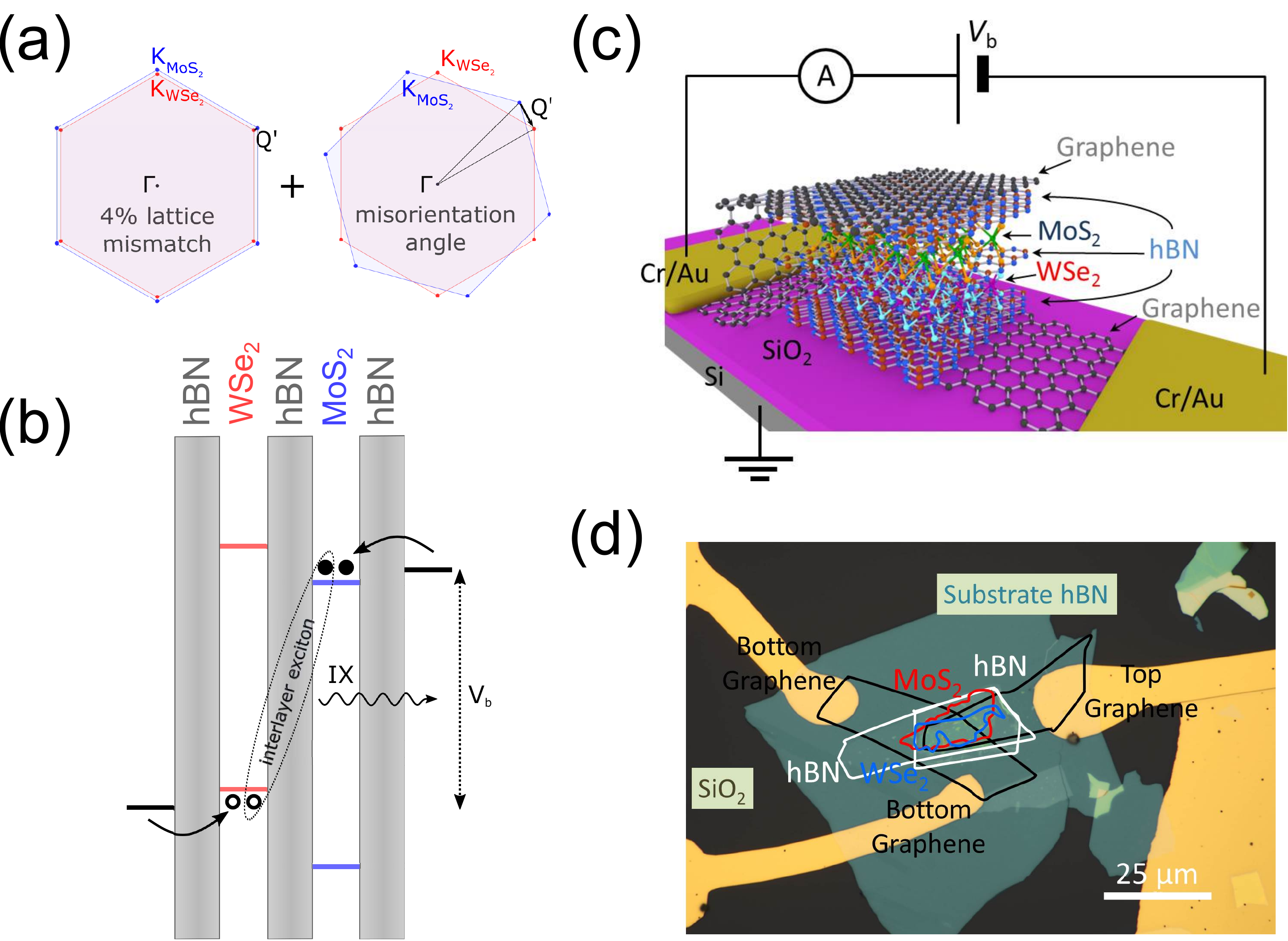}
	\caption{Sample structure and selective charge carrier injection. (a) Brillouin zones of WSe$_2$ (red) and MoS$_2$ (blue) illustrating the momentum Q' arising due to lattice mismatch and misorientation angle. (b) Schematic illustration of the type-II band alignment for the MoS$_2$/WSe$_2$ heterostructures with a middle monolayer hBN spacer. The conduction and valence band of WSe$_2$ (MoS$_2$) are represented by red (blue) lines. The hBN layers are represented by grey-shaded rectangles. The black lines depict the quasi Fermi levels in the bottom and top graphene electrodes for an applied voltage above the threshold for hole tunneling into WSe$_2$ and electron tunneling into MoS$_2$. The dashed ellipse indicates the formation of an interlayer exciton consisting of an electron in the conduction band of MoS$_2$ and a hole in the valence band of WSe$_2$. (c) Schematic drawing of the heterostructure shown in (b). (d) Optical microscope image of the active area of device B1.}
	\label{fig:fig1}
\end{figure}

In this work we employ constant electric carrier injection, which is the common operating condition for LEDs, to study the Auger recombination and exciton-exciton interactions. For this approach, without strong pumping, one has to render the radiative recombination inefficient to achieve large charge carrier densities. To this end we chose to study van der Waals (vdW) heterostructures with MoS$_2$/WSe$_2$, which are stable semiconducting TMDs that feature a type-II band alignment and a large lattice mismatch ($ \sim 4 \%$), see Figure \ref{fig:fig1} (a) and (b). Such heterostructures allow us to study long-lived interlayer excitons (IX) in the energy range of $1-1.3$~eV, important for silicon photonics and telecommunication, with an emission wavelength that can be tuned by an applied electric field. To even further suppress the radiative recombination across the indirect bandgap we produced devices with a monolayer hBN spacer between the TMDs. In such devices the Auger processes are strongly enhanced, dominating the exciton dynamics. Comparison of the behavior of samples with different spacers allows us to establish the qualitative characteristics of exciton-exciton interaction. Besides Auger processes, there is additional interest in the regime of high carrier densities motivated by recent progress in TMD based excitonic devices \cite{Unuchek2018} and theoretical predictions that vdW heterostructures could allow observing effects like the condensation of excitons \cite{Wu2015} or high-temperature super-fluidity \cite{Fogler2014}.

Another ingredient besides the reduction of radiative recombination needed to establish large IX populations is the possibility to selectively inject electrons only into one and holes only into the other material. We met this prerequisite by using tunneling injection through graphene electrodes in vertical heterostructures. In contrary to optical injection \cite{Rivera2015,Rivera2016,Nagler2017,Miller2017,Zhu2017,Kunstmann2018,Unuchek2018}, for which electrons and holes are almost exclusively created in the same and not in different materials, the electrical injection scheme circumvents the competing short-lived direct intralayer recombination, since no charge carriers of different type are present in the same material. The injection process is illustrated in Figure \ref{fig:fig1} (b) and electroluminescence (EL) becomes the method of choice to characterize the devices. Strikingly, besides the observation of the EL of the IX, we observe a large upconversion effect enabling intralayer emission of WSe$_2$ and MoS$_2$ at voltages well below the voltages corresponding to the respective excitonic bandgaps.

\section*{Results}
\subsection*{Van der Waals heterostructures for interlayer excitons}

A schematic drawing of a typical device is presented in Figure \ref{fig:fig1} (c). Si/SiO$_2$ is used as the substrate for the vdW heterostructure with the following layer sequence: Gr~/~3-5~hBN~/~1~WSe$_2$~/~0-1~hBN~/~1~MoS$_2$/ 3-5~hBN / Gr. For all devices discussed in the text the outer hBN barriers were three to five layers thick. The middle thin hBN spacer was either one layer thick or completely absent. Based on the presence of this hBN spacer we can divide the devices studied into two groups: with and without a monolayer hBN spacer separating the TMDs. Our study comprised seven devices out of which five were fabricated with and two without a monolayer hBN spacer. The devices with hBN spacer are numbered A1 - A5 and samples without hBN spacer B1, B2. For all these devices the two TMDs were aligned under the optical microscope (estimated accuracy $2^\circ$). An optical microscope image of such a sample is shown in Figure \ref{fig:fig1} (d). Details about the device fabrication and the basic optical properties can be found in ref. \onlinecite{Withers2015} and in the methods section. Figure \ref{fig:fig1} (b) illustrates the type-II band alignment of a WSe$_2$/MoS$_2$ vdW heterostructure with a monolayer hBN spacer and the tunneling pathways for charge carriers upon application of a voltage between the top and bottom graphene electrodes. The reported values for the band offsets of monolayer WSe$_2$ and MoS$_2$ are in the range of $\sim 0.6-0.7$~eV \cite{Kang2013,Gong2013,Ponomarev2018} for the conduction band offset, between $\sim 0.8-1.1$~eV \cite{Kang2013,Gong2013,Chiu2015} for the valence band offset and in the range of $\sim 0.9-1.3$~eV \cite{Kang2013,Gong2013,Ponomarev2018} for the interlayer bandgap. For the case depicted in the sketch, the voltage is large enough to enable electron injection into the MoS$_2$ conduction band and hole injection into the WSe$_2$ valence band (V$_{\text{b}} > 1$~V), but the voltage is below the threshold for direct WSe$_2$ and MoS$_2$ electron-hole injection (V$_{\text{b}} < 1.7$~V and $1.9$~V, respectively) \cite{Binder2017}. This selective injection of a given carrier type into only one material together with the large band offsets results in an extremely large charge build up at the interface and hence facilitates large IX populations.

\subsection*{Tunable electroluminescence of interlayer excitons}
\begin{figure}
	\centering
		\includegraphics[width=1.00\textwidth]{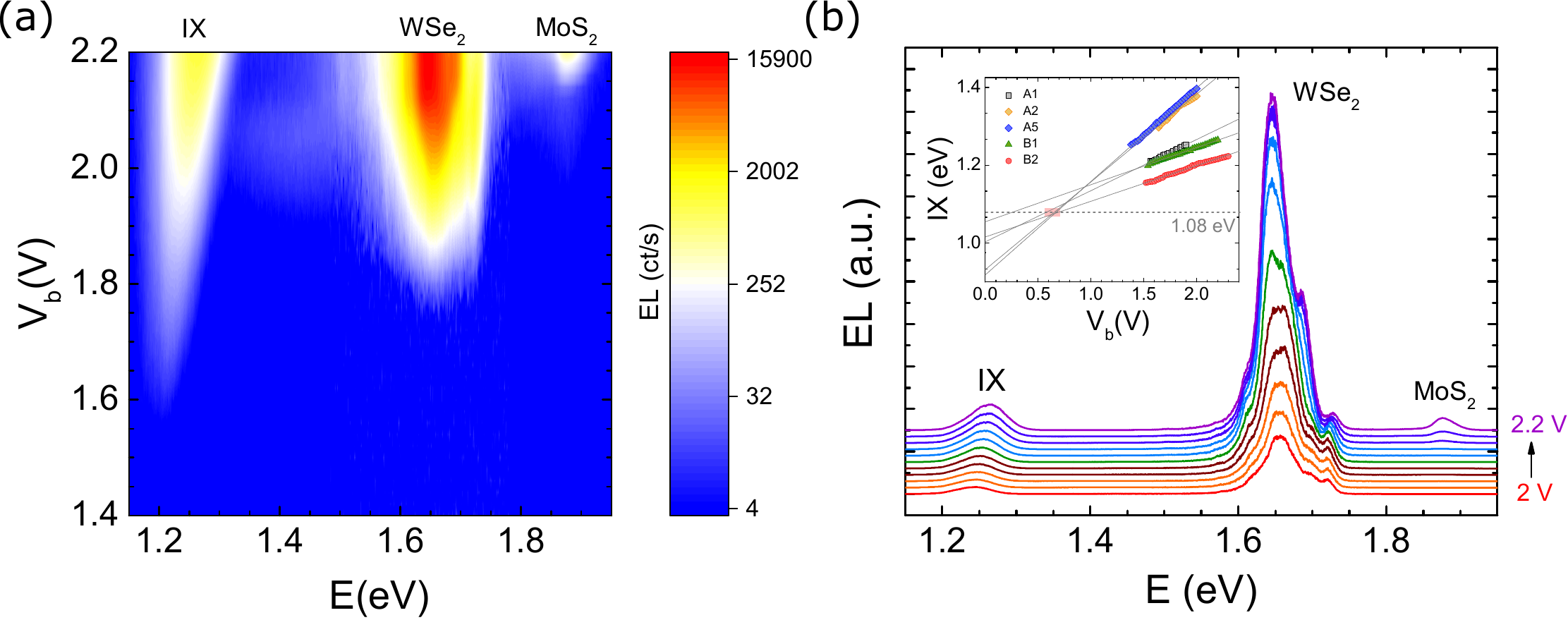}
	\caption{Interlayer excitons. (a) False color contour plot of the EL spectra as a function of bias voltage for sample B1 without monolayer hBN spacer. (b) EL spectra for biases in the range of V$_{\text{b}}$=2.0 - 2.2~V extracted from (a). The spectra are vertically shifted for clarity. The inset in panel (b) shows the peak position of the IX as a function of bias for five different samples and linear fits to the dependencies. The gray dashed line marks the energy of 1.08~eV, which is an estimation for the energy of a presumptive IX without electrically injected carriers. We obtain this value by using threshold voltages of $\sim 0.6$~V$-0.7$~V (red shaded area) for measurable carrier injection extracted from photoluminescence and reflectance contrast measurements (see Supplementary Note 4).}
	\label{fig:Fig2}
\end{figure}

First, we discuss the experimental results typical for the devices (B1 and B2) without hBN spacer between the TMDs. Figure \ref{fig:Fig2} (a) shows the evolution of the EL as a function of bias voltage in a broad energy range for sample B1. The EL spectra, Figure \ref{fig:Fig2} (b), show three contributions: an emission originating from MoS$_2$ at around 1.9~eV, a broad band at 1.7~eV related to WSe$_2$ and a third peak at around 1.2-1.3~eV, which we attribute to IX emission. We can pinpoint this peak to originate from IX since we observe a strong blueshift as a function of bias voltage. This blueshift is a consequence of the electric field that builds up in our vertical tunneling structure upon applying the bias voltage, which results in an increase of the distance between the conduction band edge of MoS$_2$ and the valence band edge of WSe$_2$ (see Figure \ref{fig:fig1} (b)). This behavior was universal for all devices studied showing linear shifts with slopes in the range of 90-200 meV/V (see Figure \ref{fig:Fig2} (b)). The different slopes are the result of the different effective thicknesses of the barriers. The shift of the IX is a measure of the electric field and can be used to estimate the IX density (see Supplementary Note 4). To determine the total IX shift one has to estimate the energy of a presumptive IX without electrically injected carriers. To this end, photoluminescence and reflectance contrast measurements (see Supplementary Note 4) were used to extract the threshold voltages for carrier injection (red shaded area in Figure \ref{fig:Fig2} (b)). The extrapolation of the linear fits to this voltage yields an energy of about 1.08~eV for a presumptive IX without electrically injected carriers, in good agreement with an interface bandgap of 1.08~eV, recently extracted from transport measurements \cite{Ponomarev2018}. The IX emission energy in vertical vdW heterostructures is easily tunable, which is an interesting feature for many prospective applications. In contrast, no significant shift of the intralayer transitions as a function of bias voltage can be observed, in agreement with the fact that transitions in the same material are not sensitive to relative band movements caused by the electric field. 

For both devices without an hBN spacer (results for device B2 are shown in Supplementary Note 3) the different contributions emerge one after another with increasing bias voltage. The IX emission appears first (V$_{\text{b}} \sim 1.5$~V), as the difference between the conduction band of MoS$_2$ and the valence band of WSe$_2$ constitutes the lowest barrier for electron-hole injection (see Figure \ref{fig:fig1} (b)).  At larger voltages (V$_{\text{b}} \sim 1.75$~V) the WSe$_2$ emission emerges, which is a result of tunneling into the intralayer excitonic states. Finally, at (V$_{\text{b}} \sim 2.1$~V) MoS$_2$ emission becomes observable in accordance to the larger band-gap.  It is interesting to note that the voltages for which EL can be observed correspond to the exciton emission energy rather than to the single particle band gap in agreement to what has been observed for vdW heterostructures with a single WSe$_2$ monolayer \cite{Binder2017}.

\subsection*{Upconverted emission of intralayer excitons}
The situation is different for the devices with a monolayer hBN spacer for which one expects a further reduced radiative recombination. The results are exemplary shown for device A1 in Figure \ref{fig:Fig3} (a) (data for other devices is presented in Supplementary Note 2). Strikingly, one observes emission from both MoS$_2$ and WSe$_2$ at bias voltages as low as 1.3~V. The emitted photons at that voltage have an energy of around 1.7~eV (WSe$_2$) and 1.9~eV (MoS$_2$), which constitutes a remarkable upconversion of $\sim$ 0.6~eV. These measurements reveal electrically driven upconversion for light-emitting devices based on vdW heterostructures. 

\begin{figure}
	\centering
		\includegraphics[width=1.00\textwidth]{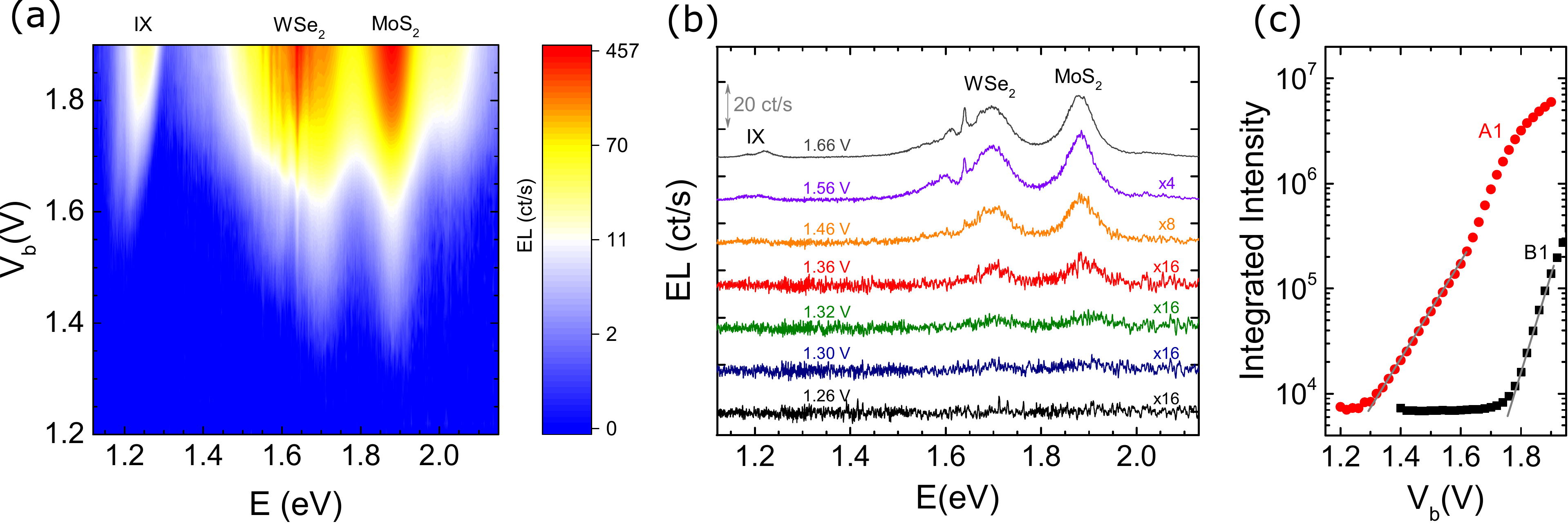}
	\caption{Upconverted electroluminescence. (a) False color contour plot of the EL spectra as a function of bias voltage for sample A1 with a monolayer hBN spacer. (b) EL spectra for seven different bias voltages extracted from (a). The spectra are vertically shifted for clarity. For a voltage of V$_{\text{b}}$=1.32~V emission at energies up to around 1.9~eV are observed, clearly illustrating the large upconversion effect. (c) Comparison of the integrated EL intensity in the spectral range of intralayer emission (1.32 - 2.37~eV) as a function of bias voltage for sample A1 (red circles) and B1 (black squares). A background signal from the response at V$_{\text{b}}$=0~V was subtracted for each spectrum. The integrated EL intensity at voltages below the onset of observable emission corresponds to the noise level of our setup of around 2ct/s per pixel (integrated over about 4000 pixels).}
	\label{fig:Fig3}
\end{figure}

An additional interesting observation is that the onset of emission is virtually the same for both MoS$_2$ and WSe$_2$. Moreover, we note that the emission from WSe$_2$ and MoS$_2$ emerges at lower applied bias voltages than the emission from the IX in contrast to samples without an hBN spacer (compare to Figure \ref{fig:Fig2} (a)). The EL spectra presented in Figure \ref{fig:Fig3} (b) also clearly show that for the sample with an hBN spacer the contribution of the intralayer excitonic emission of WSe$_2$ and MoS$_2$ is much stronger than the IX emission. These major observations allow drawing the following important conclusions. First, the process seems not to be dependent on the upconversion energy, since it shows equal onsets and intensities for both WSe$_2$ and MoS$_2$. Second, the upconversion process must be almost equally probable for electrons and holes, since in order to observe these intralayer excitons one has to lift holes into the valence band of MoS$_2$ and electrons into the conduction band of WSe$_2$ (compare Figure \ref{fig:fig1} (b)). Third, although the IX has a lower emission energy of 1.2-1.3~eV compared to around 1.9~eV (1.7~eV) for the intralayer excitons of MoS$_2$ (WSe$_2$), these intralayer excitons emerge at lower applied bias voltages in the EL spectra. Energetic upconversion was reported in literature for different inorganic semiconducting \cite{Jones2015,Manca2017,Seidel1994,Baumgartner2008,Potemski1991} as well as organic \cite{Pandey2007,Qian2010,He2016} materials. In the case of upconversion beyond the energy-scale of phonons commonly Auger-processes were identified to be responsible for the effects. The observation of substantially different onset voltages for measurable intralayer emission in samples A1 and B1 is highlighted in Figure \ref{fig:Fig3} (c) (see also Supplementary Figure 8). Clearly, the intralayer emission appears in our spectra at bias voltages as low as V$_{\text{b}} \sim 1.3$~V for sample A1, whereas this emission becomes visible at larger voltages V$_{\text{b}} \sim 1.75$~V in sample B1. One must however note that the definition of these onset voltages is somewhat arbitrary, defined by the actual experimental conditions/sensitivity (the background signal). As discussed below and illustrated in Figure \ref{fig:Fig3} (c), one should not expect a complete disappearance of Auger processes in sample B1, but their efficiency being orders of magnitude smaller in sample B1 as compared to sample A1.

\subsection*{Auger processes and upconversion mechanism}
Thanks to the purely electrical carrier injection in our devices we can rule out nonlinear effects involving photon absorption. It is therefore intuitive to investigate whether Auger processes can account for the above described observations. Such an Auger process would proceed in the following manner: (i) An electron in MoS$_2$ and a hole in WSe$_2$ form an IX that recombines non-radiatively. (ii) The excess energy is transferred to another electron in the MoS$_2$ conduction band. (iii) Since the transferred energy is larger than the band offset the electron can tunnel into the WSe$_2$ conduction band. (iv) Due to the large number of holes present in the valence band of WSe$_2$, the electron can form a WSe$_2$ intralayer exciton and recombine radiatively, which gives rise to the characteristic light emission for WSe$_2$ monolayers. A similar process would be possible for holes in the valence band of WSe$_2$. With such a process one can account for the energy independence of the process, which is true as long as the energy of the indirect transition is larger than the band offsets. However, the second aspect identified above, i.e. same probability for the upconversion of electrons and holes, is hard to imagine in such a case. For such an Auger process to be efficient there should be a state to which the charge carrier can be excited. Yet in the case of WSe$_2$ there are no bands with an energy distance of the order of the IX energy in the band structure around the K point to which a hole could be excited \cite{Danovich2016}. Even if one includes larger momenta it is hard to imagine that such a process would be equally probable as compared to electrons for which such bands should be present. In this picture it is also difficult to explain why upconversion sets in at lower bias voltages than the IX emission. In order to overcome this conceptual discrepancies, we propose an excitonic Auger process instead of the above described single particle considerations in agreement with the exciton-exciton interactions observed in optical pump-probe experiments \cite{Sun2014,Poellmann2015,Yu2016,Aivazian2017,Kulig2018}.

\begin{figure}
	\centering
		\includegraphics[width=0.65\textwidth]{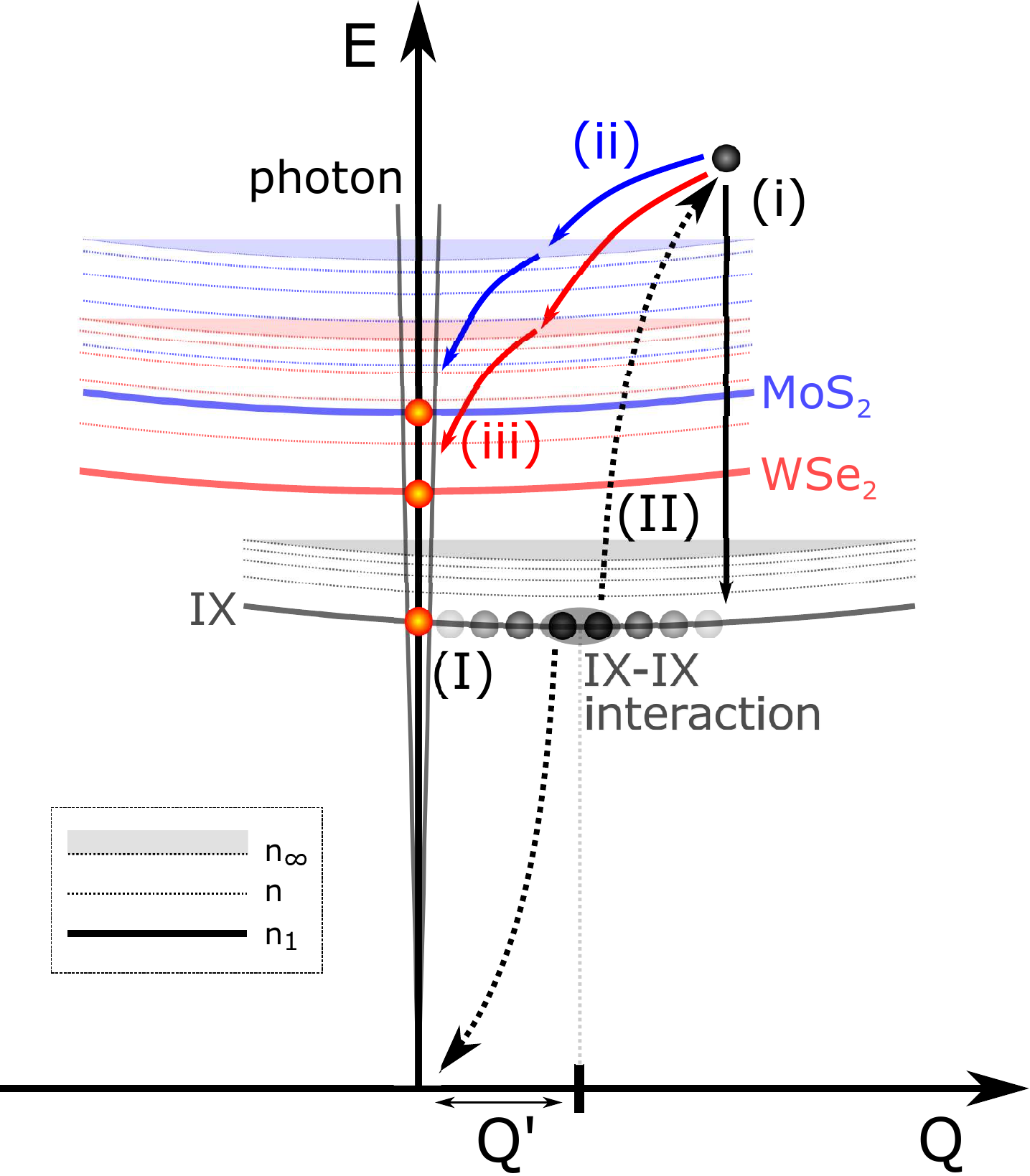}
	\caption{Mechanism of upconverted emission in the two-particle picture. The solid lines represent the excitonic ground state dispersion n$_{\text{1}}$ of the IX (grey), MoS$_2$ (blue) and WSe$_2$ (red). The circles stand for excitons. Q$^{\prime}$ is the momentum mismatch as defined in Figure \ref{fig:fig1} (a). The dashed lines indicate excited states n and the shaded area marks the excitonic continuum n$_{\infty}$. The grey-scale shading of circles schematically pictures the momentum distribution of excitons. A bright shading indicates less excitons for a given momentum than a dark shading. The photon dispersion is overlaid (grey lines) to mark the region of effective radiative recombination (orange circles). For the situation depicted, the bias voltage is below the threshold for direct intralayer charge injection. Mechanism (I) illustrates radiative IX emission facilitated by an increasing number of IX with large momenta. Mechanism (II) depicts excitonic Auger processes. The grey ellipse schematically highlights the interaction between two exemplary excitons. As a result of the interaction one exciton recombines non-radiatively and transfers the energy to the other exciton (arrows with dotted lines). Relaxation: (i) describes relaxation back to the IX ground state (exciton-exciton annihilation), (ii) and (iii) relaxation to MoS$_2$ and WSe$_2$ respectively, which leads to upconverted intralayer emission.}
	\label{fig:X_Sketch}
\end{figure}

Figure \ref{fig:X_Sketch} presents the proposed mechanism responsible for the upconverted EL in the two-particle picture. The grey (red, blue) solid parabola represents the exciton dispersion of the IX (WSe$_2$, MoS$_2$). The dashed lines represent exited states and the shaded filled area stands for the excitonic continuum. Due to the lattice mismatch, the parabola describing the IX is shifted towards higher momenta. If there is an additional misalignment angle this will further increase the distance Q' (see Figure \ref{fig:fig1} (a)). As a result of the effective selective injection of carriers the population of IX increases. The distribution function of the population is schematically depicted by the shading of the grey circles. The orange circles stand for excitons that possess almost zero center of mass velocity and hence can recombine radiatively (light cones). 

In this picture, the IX have to compensate for the additional momentum Q' first in order to recombine radiatively. This can be achieved in two ways: (I) by increasing the IX population at larger voltages. Since with increasing bias electrons and holes that form the IX possess larger and larger momenta, a significant amount of the IX population will as well extend more and more towards larger momenta. The momentum range of efficient radiative recombination is determined by the photon dispersion, as indicated in Figure \ref{fig:X_Sketch}. If their momentum roughly equals Q', the IX can recombine radiatively.  Otherwise, the IX remains optically silent and gives rise to an increasing population that enables mechanism (II) through IX-IX collisions. As a result, one IX recombines non-radiatively and transfers the energy and momentum to a second IX (dotted arrows in Figure \ref{fig:X_Sketch}). This IX is excited into the continuum of excitonic states and has now three different decay channels: (i) the IX relaxes back into the IX ground state. This scenario is also called exciton-exciton annihilation. (ii) corresponds to the case where the now almost delocalized exciton relaxes into a state that corresponds to the MoS$_2$ intralayer exciton. Please note that the relaxation process shown here is indirect in real space, i.e. that in the single particle picture this would correspond to a tunneling of the electron into the MoS$_2$ conduction band. Similarly possible is (iii) for which the excited exciton relaxes to a WSe$_2$ exciton state. For both (ii) and (iii), the relaxation continues via intralayer excitonic states until the exciton reaches the bottom of the dispersion relation, where it can recombine radiatively. 

Within this picture we can imagine (ii) and (iii) to be equally probable, since there is a plethora of excited excitonic states for both materials, in contrast to the lack of higher energetic hole bands around the K point in the single particle picture. One can also explain the lower voltages for the onset of upconversion compared to the onset for the IX emission. In order to recombine radiatively the IX have to compensate for the momentum Q' (I), but below this threshold there is already a significant population of optically silent IX present that can give rise to the upconversion (II). This difference in onsets should strongly depend on the misorientation angle that adds to the momentum Q' due to lattice mismatch. Based on literature values for the effective masses and lattice parameters \cite{Kormanyos2015} and by using an estimation of the momentum mismatch of $0.04 \cdot \overline{\Gamma K}$ for zero misalignment and $0.1 \cdot \overline{\Gamma K}$  for $5^{\circ}$ misalignment \cite{Yu2015} we can estimate the additional kinetic energy needed to recombine radiatively to be $\sim 10$~meV and $\sim 60$~meV, respectively. These energies indicate that such an emission could be temperature activated, but depends sensitively on the misalignment angle. It has been shown theoretically that for such structures one should expect six light cones irrespective of the layer misalignment angle, which are situated at a non-zero center of mass velocity \cite{Yu2015}. Consequently, the IX have to posses a large kinetic momentum or have to be scattered by defects or phonons into the light cones to recombine radiatively. Otherwise these IX remain dark as shown in Figure \ref{fig:X_Sketch}. This fact renders the observation of PL of the IX very ineffective \cite{Zhu2017,Kunstmann2018,Ponomarev2018}, but on the other hand offers an ideal platform to study many-body interactions allowing more easily reaching the large IX population regime. In the more widely studied system of MoSe$_2$/WSe$_2$ the lattice mismatch is small ($<0.1 \%$) which makes the PL of IX observable allowing to study the properties of the IX more directly \cite{Rivera2015,Rivera2016,Ross2017,Nagler2017,Miller2017}, but consequently not allowing for large IX populations. For our samples measured at low temperatures no PL of the IX could be observed in agreement with other recent reports \cite{Zhu2017,Kunstmann2018,Ponomarev2018} even with additional applied voltages. Please note that we only consider $K-K$ interlayer excitons where the electron and hole stem from the K points of the respective TMD layers. Recently, $\Gamma - K$ interlayer excitons were observed and identified in the photoluminescence of MoS$_2$/WSe$_2$ vdW heterostructures\cite{Kunstmann2018}. These $\Gamma - K$ interlayer excitons appear at larger energies of about 1.6~eV, which agrees with other reports of interlayer excitons in this material system \cite{Fang2014,Chiu2014,Unuchek2018}. In this work we were able to reveal the emission spectra of $K-K$ interlayer excitons which are not observable in PL. This observation is enabled thanks to our selective electrical injection mechanism, leading to large IX populations.

\section*{Discussion}
Out of the seven devices studied, three showed upconverted EL emission. An additional device showed similar behavior -- e.g. intralayer emission before IX -- but at larger voltages suggesting that not the whole applied voltage dropped across the active region of the vdW heterostructure (see more detailed information on every device in Supplementary Note 1). All of these devices feature a monolayer hBN spacer. The two devices without hBN spacer did not show upconversion, as well as one sample with hBN spacer for which, however, charge injection was low and asymmetric. In our picture, we can also account for the behavior of the devices that do not show upconversion. Here, the consecutive appearance of different contributions with increasing bias voltage (compare Figure \ref{fig:Fig2} (a)) is reflected by the increasing energetic positions of bright excitons around zero momentum (see Figure \ref{fig:X_Sketch}) into which the charge carriers can tunnel directly at larger voltages. 

Our results suggest that the probability for Auger processes and upconversion versus radiative IX recombination can be effectively tuned by changing the thickness of the middle hBN barrier. The presence of an hBN spacer between the TMDs strongly influences the radiative recombination rate for IX emission in the vicinity of the light cone. By introducing an hBN spacer we increase the distance between the electron and the hole and consequently decrease the wave-function overlap. We can therefore expect the radiative recombination rate to be quenched in the case of devices from group A. Indeed, we observe that the upconverted intralayer emission is more intense than the IX emission (see Figure \ref{fig:Fig3}). Devices with an hBN spacer hence favor larger IX densities, since the radiative IX recombination at almost zero momentum becomes less effective.

To experimentally confirm this conjecture we estimated the IX density with two different methods: (i) by analyzing the apparent evolution of the exciton versus trion resonances in the MoS$_2$ layer in photoluminescence (PL) and reflectance contrast (RC) measured as a function of bias voltage and (ii) by applying a simple capacitor model to account for the observed blueshift of the IX peak in the EL measured as a function of applied bias voltage, thus deducing the built-in electron-hole charge. Both methods (see Supplementary Note 4 for details on measurements and analysis) yield IX densities in the range of $10^{12}$~cm$^{-2}$ for sample A1, but distinctly smaller densities, in the range of $10^{11}$~cm$^{-2}$, for sample B1. With these findings we indeed confirm our expectations that a larger accumulation of interlayer excitons appears in samples with an hBN spacer, favoring the observation of Auger processes in such structures. 

However, the presence or absence (strong or weak efficiency) of the Auger processes should also be discussed with respect to an apparent ratio d$/$a$_{\text{B}}$  of the inter exciton distance d to the exciton Bohr radius a$_{\text{B}}$. With n $= 3.7 \cdot 10^{12}$~cm$^{-2}$ for sample A1 and  n $=6.1 \cdot 10^{11}$~cm$^{-2}$  for sample B1 (see Supplementary Note 4) we obtain d$_{\text{A1}}=2 /\sqrt{\pi n} \sim 5.9$~nm and d$_{\text{B1}} \sim 14.4$~nm, respectively for samples A1 and B1. To estimate a$_{\text{B}}$ we note that the Bohr radius of an interlayer exciton is expected to be about a factor of two larger than that of the intralayer exciton \cite{Latini2017}. The exciton Bohr radius for intralayer  excitons of semiconducting monolayer TMDs is known to be sensitive to the dielectric surrounding and varies in the range of 1 - 1.7~nm \cite{Wang2018,Stier2018}. Thus, given their hBN encapsulation, we roughly estimate a$_{\text{B}} \sim 3$~nm for the interlayer exciton Bohr radius in our structures and conclude that d$_{\text{A1}}/$a$_{\text{B}} \sim 2$ for sample A1 and a significantly larger value d$_{\text{B1}}/$a$_{\text{B}} \sim 5$ for sample B1. The conditions met by sample A1 are close to the Mott transition for interlayer excitons, which enables the observation of many body effects in this case \cite{Wang2018}. In addition, one may speculate that d$/$a$_{\text{B}}$ can be further reduced in sample A1, as the introduction of an hBN spacer could lead to the reduction of the binding energy of the intralayer exciton in sample A1, as compared to sample B1. However, this reduction was calculated to be rather small \cite{Latini2017}, thus charge accumulation seems to be a decisive factor for the observation of the Auger-type processes in our samples.

An expected super-linear (ideally quadratic) behavior of the intensity of the Auger upconverted emission as a function of the IX density is another issue to be considered. It is tempting to trace this behavior by investigating the dependence of the emission intensity versus driving current. However, this dependence may also be super-linear due to trivial effects not related to the Auger process, like leakage currents due to device imperfections or exciton trapping centers, which are also observed in III-V quantum well structures \cite{Combescot2017}. For most of our devices we obtain an overestimation of the super-linear behavior of emission intensity as a function of driving current, due to current contributions that are not related to the emission process. Instead, for device A1, showing excellent I-V characteristics, the extracted super-linear trend in the regime of the Auger upconverted emission appears reliable, since in the case of this device the trend becomes practically linear in the direct intralayer injection regime, when the applied voltages exceed the excitonic bandgaps of the TMD monolayers (see Supplementary Figure 9). It is clear that more work on the role and nature of defects and a better control of the interfaces in vdW heterostructures is needed to tackle this issue for all devices.

In conclusion, we studied Auger processes in especially chosen type-II vdW heterostructures that enable large populations of optically silent IX. The double indirect nature of these excitons allows recovering a part of the excited Auger carriers through relaxation in optically active states. These states lie higher in energy than the initial IX, which results in the emission of upconverted photons. The large IX populations established thanks to the purely electrical tunnel injection allowed us to observe EL of otherwise silent IX, which shifts due to the electric field. It is shown that the IX energy can be tuned by up to 200~meV/V. A purely excitonic Auger process is proposed that, in contrary to Auger processes based on the single-particle picture, accounts for the major characteristics of the EL as a function of bias voltage. The results suggest that the efficiency of the upconversion mechanism depends on the hBN layer between the active TMDs. The revealed variable nature of the IX-IX interactions is of key importance for future TMD based optoelectronic device engineering and is important for any attempt aiming towards exciton condensation or super-fluidity in such structures. 

\section*{Methods}

\subsection*{Device Fabrication}
All devices were fabricated on doped silicon substrates covered with silicon oxide 90~nm thick. Top graphene flakes were exfoliated on a silicon substrate spin coated with PMGI (polymethylglutarimide) and PMMA (polymethyl methacrylate). After dissolving PMGI, PMMA membranes with graphene were used to pick up other flakes. MoS$_2$ and WSe$_2$ were exfoliated on a silicon oxide substrate (290~nm thick) spin coated with PPC (polypropylene carbonate). Thin boron nitride and bottom graphene were exfoliated on silicon oxide substrates 70 and 90~nm thick, respectively. In some devices, we used substrate hBN about 30~nm thick exfoliated on silica substrates 90~nm thick.
Light-emitting diodes were assembled by picking TMDs and hBN flakes one after another with the top graphene flake and transferring the resulting stack on top of bottom graphene. The PMMA membrane was then dissolved in acetone. In some devices, the flakes were picked up by top graphene and peeled onto the substrate hBN. After the assembly, electron-beam lithography was used to define a mask for contacts followed by evaporation of Chromium/Gold (Cr/Au, 3/50nm).

\subsection*{Optoelectronic measurements}
The optoelectronic measurements were performed using two different helium flow cryostats (Janis ST-500, Oxford Instruments MicrostatHires) and a helium bath cryostat. All the measurements shown in the main text were taken at liquid helium temperature. The signal was collected using a 0.5~m long spectrometer equipped with liquid nitrogen cooled charge-couple-device (CCD) camera. Electrical measurements were performed using a Keithley 2450 source-measure unit synchronized with the spectrometer. A diode type behavior was observed for all devices.

\subsection*{Data availability}
The data supporting the findings of this work are available from the corresponding author upon reasonable request.

\bibliographystyle{naturemag} 
\bibliography{references}

\begin{thebibliography}{10}
\expandafter\ifx\csname url\endcsname\relax
  \def\url#1{\texttt{#1}}\fi
\expandafter\ifx\csname urlprefix\endcsname\relax\def\urlprefix{URL }\fi
\providecommand{\bibinfo}[2]{#2}
\providecommand{\eprint}[2][]{\url{#2}}

\bibitem{Pandey2007}
\bibinfo{author}{Pandey, A.~K.} \& \bibinfo{author}{Nunzi, J.-M.}
\newblock \bibinfo{title}{Upconversion injection in
  rubrene/perylene-diimide-heterostructure electroluminescent diodes}.
\newblock \emph{\bibinfo{journal}{Appl. Phys. Lett.}}
  \textbf{\bibinfo{volume}{90}}, \bibinfo{pages}{263508}
  (\bibinfo{year}{2007}).

\bibitem{Qian2010}
\bibinfo{author}{Qian, L.} \emph{et~al.}
\newblock \bibinfo{title}{Electroluminescence from light-emitting polymer/{ZnO}
  nanoparticle heterojunctions at sub-bandgap voltages}.
\newblock \emph{\bibinfo{journal}{Nano Today}} \textbf{\bibinfo{volume}{5}},
  \bibinfo{pages}{384--389} (\bibinfo{year}{2010}).

\bibitem{He2016}
\bibinfo{author}{He, S.-J.}, \bibinfo{author}{Wang, D.-K.},
  \bibinfo{author}{Jiang, N.}, \bibinfo{author}{Tse, J.~S.} \&
  \bibinfo{author}{Lu, Z.-H.}
\newblock \bibinfo{title}{Tunable excitonic processes at organic
  heterojunctions}.
\newblock \emph{\bibinfo{journal}{Adv. Mater.}} \textbf{\bibinfo{volume}{28}},
  \bibinfo{pages}{649--654} (\bibinfo{year}{2016}).

\bibitem{Iveland2013}
\bibinfo{author}{Iveland, J.}, \bibinfo{author}{Martinelli, L.},
  \bibinfo{author}{Peretti, J.}, \bibinfo{author}{Speck, J.~S.} \&
  \bibinfo{author}{Weisbuch, C.}
\newblock \bibinfo{title}{Direct measurement of {A}uger electrons emitted from
  a semiconductor light-emitting diode under electrical injection:
  identification of the dominant mechanism for efficiency droop}.
\newblock \emph{\bibinfo{journal}{Phys. Rev. Lett.}}
  \textbf{\bibinfo{volume}{110}}, \bibinfo{pages}{177406}
  (\bibinfo{year}{2013}).

\bibitem{Sun2014}
\bibinfo{author}{Sun, D.} \emph{et~al.}
\newblock \bibinfo{title}{Observation of rapid exciton--exciton annihilation in
  monolayer molybdenum disulfide}.
\newblock \emph{\bibinfo{journal}{Nano Lett.}} \textbf{\bibinfo{volume}{14}},
  \bibinfo{pages}{5625--5629} (\bibinfo{year}{2014}).

\bibitem{Yuan2015}
\bibinfo{author}{Yuan, L.} \& \bibinfo{author}{Huang, L.}
\newblock \bibinfo{title}{Exciton dynamics and annihilation in {WS}$_2$ {2D}
  semiconductors}.
\newblock \emph{\bibinfo{journal}{Nanoscale}} \textbf{\bibinfo{volume}{7}},
  \bibinfo{pages}{7402--7408} (\bibinfo{year}{2015}).

\bibitem{Poellmann2015}
\bibinfo{author}{P{\"o}llmann, C.} \emph{et~al.}
\newblock \bibinfo{title}{Resonant internal quantum transitions and femtosecond
  radiative decay of excitons in monolayer {WSe}$_2$}.
\newblock \emph{\bibinfo{journal}{Nat. Mat.}} \textbf{\bibinfo{volume}{14}},
  \bibinfo{pages}{889} (\bibinfo{year}{2015}).

\bibitem{Yu2016}
\bibinfo{author}{Yu, Y.} \emph{et~al.}
\newblock \bibinfo{title}{Fundamental limits of exciton-exciton annihilation
  for light emission in transition metal dichalcogenide monolayers}.
\newblock \emph{\bibinfo{journal}{Phys. Rev. B}} \textbf{\bibinfo{volume}{93}},
  \bibinfo{pages}{201111} (\bibinfo{year}{2016}).

\bibitem{Aivazian2017}
\bibinfo{author}{Aivazian, G.} \emph{et~al.}
\newblock \bibinfo{title}{Many-body effects in nonlinear optical responses of
  {2D} layered semiconductors}.
\newblock \emph{\bibinfo{journal}{2D Mater.}} \textbf{\bibinfo{volume}{4}},
  \bibinfo{pages}{025024} (\bibinfo{year}{2017}).

\bibitem{Kulig2018}
\bibinfo{author}{Kulig, M.} \emph{et~al.}
\newblock \bibinfo{title}{Exciton diffusion and halo effects in monolayer
  semiconductors}.
\newblock \emph{\bibinfo{journal}{Phys. Rev. Lett.}}
  \textbf{\bibinfo{volume}{120}}, \bibinfo{pages}{207401}
  (\bibinfo{year}{2018}).

\bibitem{Unuchek2018}
\bibinfo{author}{Unuchek, D.} \emph{et~al.}
\newblock \bibinfo{title}{Room-temperature electrical control of exciton flux
  in a van der {Waals} heterostructure.}
\newblock \emph{\bibinfo{journal}{Nature}} \textbf{\bibinfo{volume}{560}},
  \bibinfo{pages}{340--344} (\bibinfo{year}{2018}).

\bibitem{Wu2015}
\bibinfo{author}{Wu, F.-C.}, \bibinfo{author}{Xue, F.} \&
  \bibinfo{author}{MacDonald, A.}
\newblock \bibinfo{title}{Theory of two-dimensional spatially indirect
  equilibrium exciton condensates}.
\newblock \emph{\bibinfo{journal}{Phys. Rev. B}} \textbf{\bibinfo{volume}{92}},
  \bibinfo{pages}{165121} (\bibinfo{year}{2015}).

\bibitem{Fogler2014}
\bibinfo{author}{Fogler, M.}, \bibinfo{author}{Butov, L.} \&
  \bibinfo{author}{Novoselov, K.}
\newblock \bibinfo{title}{High-temperature superfluidity with indirect excitons
  in van der {Waals} heterostructures}.
\newblock \emph{\bibinfo{journal}{Nat. Commun.}} \textbf{\bibinfo{volume}{5}},
  \bibinfo{pages}{4555} (\bibinfo{year}{2014}).

\bibitem{Rivera2015}
\bibinfo{author}{Rivera, P.} \emph{et~al.}
\newblock \bibinfo{title}{Observation of long-lived interlayer excitons in
  monolayer {MoSe}$_2$ -- {WSe}$_2$ heterostructures}.
\newblock \emph{\bibinfo{journal}{Nat. Commun.}} \textbf{\bibinfo{volume}{6}},
  \bibinfo{pages}{6242} (\bibinfo{year}{2015}).

\bibitem{Rivera2016}
\bibinfo{author}{Rivera, P.} \emph{et~al.}
\newblock \bibinfo{title}{Valley-polarized exciton dynamics in a {2D}
  semiconductor heterostructure}.
\newblock \emph{\bibinfo{journal}{Science}} \textbf{\bibinfo{volume}{351}},
  \bibinfo{pages}{688--691} (\bibinfo{year}{2016}).

\bibitem{Nagler2017}
\bibinfo{author}{Nagler, P.} \emph{et~al.}
\newblock \bibinfo{title}{Interlayer exciton dynamics in a dichalcogenide
  monolayer heterostructure}.
\newblock \emph{\bibinfo{journal}{2D Mater.}} \textbf{\bibinfo{volume}{4}},
  \bibinfo{pages}{025112} (\bibinfo{year}{2017}).

\bibitem{Miller2017}
\bibinfo{author}{Miller, B.} \emph{et~al.}
\newblock \bibinfo{title}{Long-lived direct and indirect interlayer excitons in
  van der {Waals} heterostructures}.
\newblock \emph{\bibinfo{journal}{Nano Lett.}} \textbf{\bibinfo{volume}{17}},
  \bibinfo{pages}{5229--5237} (\bibinfo{year}{2017}).

\bibitem{Zhu2017}
\bibinfo{author}{Zhu, H.} \emph{et~al.}
\newblock \bibinfo{title}{Interfacial charge transfer circumventing momentum
  mismatch at two-dimensional van der {Waals} heterojunctions}.
\newblock \emph{\bibinfo{journal}{Nano Lett.}} \textbf{\bibinfo{volume}{17}},
  \bibinfo{pages}{3591--3598} (\bibinfo{year}{2017}).

\bibitem{Kunstmann2018}
\bibinfo{author}{Kunstmann, J.} \emph{et~al.}
\newblock \bibinfo{title}{Momentum-space indirect interlayer excitons in
  transition metal dichalcogenide van der {Waals} heterostructures}.
\newblock \emph{\bibinfo{journal}{Nat. Phys.}} \textbf{\bibinfo{volume}{14}},
  \bibinfo{pages}{801--805} (\bibinfo{year}{2018}).

\bibitem{Withers2015}
\bibinfo{author}{Withers, F.} \emph{et~al.}
\newblock \bibinfo{title}{Light-emitting diodes by band-structure engineering
  in van der {Waals} heterostructures}.
\newblock \emph{\bibinfo{journal}{Nat. Mat.}} \textbf{\bibinfo{volume}{14}},
  \bibinfo{pages}{301--306} (\bibinfo{year}{2015}).

\bibitem{Kang2013}
\bibinfo{author}{Kang, J.}, \bibinfo{author}{Tongay, S.},
  \bibinfo{author}{Zhou, J.}, \bibinfo{author}{Li, J.} \& \bibinfo{author}{Wu,
  J.}
\newblock \bibinfo{title}{Band offsets and heterostructures of two-dimensional
  semiconductors}.
\newblock \emph{\bibinfo{journal}{Appl. Phys. Lett.}}
  \textbf{\bibinfo{volume}{102}}, \bibinfo{pages}{012111}
  (\bibinfo{year}{2013}).

\bibitem{Gong2013}
\bibinfo{author}{Gong, C.} \emph{et~al.}
\newblock \bibinfo{title}{Band alignment of two-dimensional transition metal
  dichalcogenides: Application in tunnel field effect transistors}.
\newblock \emph{\bibinfo{journal}{Appl. Phys. Lett.}}
  \textbf{\bibinfo{volume}{103}}, \bibinfo{pages}{053513}
  (\bibinfo{year}{2013}).

\bibitem{Ponomarev2018}
\bibinfo{author}{Ponomarev, E.}, \bibinfo{author}{Ubrig, N.},
  \bibinfo{author}{Guti{\'e}rrez-Lezama, I.}, \bibinfo{author}{Berger, H.} \&
  \bibinfo{author}{Morpurgo, A.~F.}
\newblock \bibinfo{title}{Semiconducting van der {Waals} interfaces as
  artificial semiconductors}.
\newblock \emph{\bibinfo{journal}{Nano Lett.}} \textbf{\bibinfo{volume}{18}},
  \bibinfo{pages}{5146--5152} (\bibinfo{year}{2018}).

\bibitem{Chiu2015}
\bibinfo{author}{Chiu, M.-H.} \emph{et~al.}
\newblock \bibinfo{title}{Determination of band alignment in the single-layer
  {MoS}$_2$/{WSe}$_2$ heterojunction}.
\newblock \emph{\bibinfo{journal}{Nat. Commun.}} \textbf{\bibinfo{volume}{6}},
  \bibinfo{pages}{7666} (\bibinfo{year}{2015}).

\bibitem{Binder2017}
\bibinfo{author}{Binder, J.} \emph{et~al.}
\newblock \bibinfo{title}{Sub-bandgap voltage electroluminescence and
  magneto-oscillations in a {WSe}$_2$ light-emitting van der {Waals}
  heterostructure}.
\newblock \emph{\bibinfo{journal}{Nano Lett.}} \textbf{\bibinfo{volume}{17}},
  \bibinfo{pages}{1425--1430} (\bibinfo{year}{2017}).

\bibitem{Jones2015}
\bibinfo{author}{Jones, A.~M.} \emph{et~al.}
\newblock \bibinfo{title}{Excitonic luminescence upconversion in a
  two-dimensional semiconductor}.
\newblock \emph{\bibinfo{journal}{Nat. Phys.}} \textbf{\bibinfo{volume}{12}},
  \bibinfo{pages}{323--327} (\bibinfo{year}{2015}).

\bibitem{Manca2017}
\bibinfo{author}{Manca, M.} \emph{et~al.}
\newblock \bibinfo{title}{Enabling valley selective exciton scattering in
  monolayer {WSe}$_2$ through upconversion}.
\newblock \emph{\bibinfo{journal}{Nat. Commun.}} \textbf{\bibinfo{volume}{8}},
  \bibinfo{pages}{14927} (\bibinfo{year}{2017}).

\bibitem{Seidel1994}
\bibinfo{author}{Seidel, W.}, \bibinfo{author}{Titkov, A.},
  \bibinfo{author}{Andr{\'e}, J.}, \bibinfo{author}{Voisin, P.} \&
  \bibinfo{author}{Voos, M.}
\newblock \bibinfo{title}{High-efficiency energy up-conversion by an "{A}uger
  fountain" at an {InP-AlInAs} type-{II} heterojunction}.
\newblock \emph{\bibinfo{journal}{Phys. Rev. Lett.}}
  \textbf{\bibinfo{volume}{73}}, \bibinfo{pages}{2356} (\bibinfo{year}{1994}).

\bibitem{Baumgartner2008}
\bibinfo{author}{Baumgartner, A.}, \bibinfo{author}{Chaggar, A.},
  \bibinfo{author}{Patan{\`e}, A.}, \bibinfo{author}{Eaves, L.} \&
  \bibinfo{author}{Henini, M.}
\newblock \bibinfo{title}{Upconversion electroluminescence in {InAs} quantum
  dot light-emitting diodes}.
\newblock \emph{\bibinfo{journal}{Appl. Phys. Lett.}}
  \textbf{\bibinfo{volume}{92}}, \bibinfo{pages}{091121}
  (\bibinfo{year}{2008}).

\bibitem{Potemski1991}
\bibinfo{author}{Potemski, M.} \emph{et~al.}
\newblock \bibinfo{title}{Auger recombination within {Landau} levels in a
  two-dimensional electron gas}.
\newblock \emph{\bibinfo{journal}{Phys. Rev. Lett.}}
  \textbf{\bibinfo{volume}{66}}, \bibinfo{pages}{2239} (\bibinfo{year}{1991}).

\bibitem{Danovich2016}
\bibinfo{author}{Danovich, M.}, \bibinfo{author}{Z{\'o}lyomi, V.},
  \bibinfo{author}{Falko, V.~I.} \& \bibinfo{author}{Aleiner, I.~L.}
\newblock \bibinfo{title}{Auger recombination of dark excitons in {WS}$_2$ and
  {WSe}$_2$ monolayers}.
\newblock \emph{\bibinfo{journal}{2D Mater.}} \textbf{\bibinfo{volume}{3}},
  \bibinfo{pages}{035011} (\bibinfo{year}{2016}).

\bibitem{Kormanyos2015}
\bibinfo{author}{Korm{\'a}nyos, A.} \emph{et~al.}
\newblock \bibinfo{title}{K{\textperiodcentered} p theory for two-dimensional
  transition metal dichalcogenide semiconductors}.
\newblock \emph{\bibinfo{journal}{2D Mater.}} \textbf{\bibinfo{volume}{2}},
  \bibinfo{pages}{022001} (\bibinfo{year}{2015}).

\bibitem{Yu2015}
\bibinfo{author}{Yu, H.}, \bibinfo{author}{Wang, Y.}, \bibinfo{author}{Tong,
  Q.}, \bibinfo{author}{Xu, X.} \& \bibinfo{author}{Yao, W.}
\newblock \bibinfo{title}{Anomalous light cones and valley optical selection
  rules of interlayer excitons in twisted heterobilayers}.
\newblock \emph{\bibinfo{journal}{Phys. Rev. Lett.}}
  \textbf{\bibinfo{volume}{115}}, \bibinfo{pages}{187002}
  (\bibinfo{year}{2015}).

\bibitem{Ross2017}
\bibinfo{author}{Ross, J.~S.} \emph{et~al.}
\newblock \bibinfo{title}{Interlayer exciton optoelectronics in a {2D}
  heterostructure pn junction}.
\newblock \emph{\bibinfo{journal}{Nano Lett.}} \textbf{\bibinfo{volume}{17}},
  \bibinfo{pages}{638--643} (\bibinfo{year}{2017}).

\bibitem{Fang2014}
\bibinfo{author}{Fang, H.} \emph{et~al.}
\newblock \bibinfo{title}{Strong interlayer coupling in van der {Waals}
  heterostructures built from single-layer chalcogenides}.
\newblock \emph{\bibinfo{journal}{Proc. Natl. Acad. Sci. U.S.A.}}
  \textbf{\bibinfo{volume}{111}}, \bibinfo{pages}{6198--6202}
  (\bibinfo{year}{2014}).

\bibitem{Chiu2014}
\bibinfo{author}{Chiu, M.-H.} \emph{et~al.}
\newblock \bibinfo{title}{Spectroscopic signatures for interlayer coupling in
  {MoS}$_2$-{WSe}$_2$ van der {W}aals stacking}.
\newblock \emph{\bibinfo{journal}{ACS Nano}} \textbf{\bibinfo{volume}{8}},
  \bibinfo{pages}{9649--9656} (\bibinfo{year}{2014}).

\bibitem{Latini2017}
\bibinfo{author}{Latini, S.}, \bibinfo{author}{Winther, K.~T.},
  \bibinfo{author}{Olsen, T.} \& \bibinfo{author}{Thygesen, K.~S.}
\newblock \bibinfo{title}{Interlayer excitons and band alignment in
  {MoS}$_2$/h{BN}/{WSe}$_2$ van der {W}aals heterostructures}.
\newblock \emph{\bibinfo{journal}{Nano Lett.}} \textbf{\bibinfo{volume}{17}},
  \bibinfo{pages}{938--945} (\bibinfo{year}{2017}).

\bibitem{Wang2018}
\bibinfo{author}{Wang, G.} \emph{et~al.}
\newblock \bibinfo{title}{Colloquium: Excitons in atomically thin transition
  metal dichalcogenides}.
\newblock \emph{\bibinfo{journal}{Rev. Mod. Phys.}}
  \textbf{\bibinfo{volume}{90}}, \bibinfo{pages}{021001}
  (\bibinfo{year}{2018}).

\bibitem{Stier2018}
\bibinfo{author}{Stier, A.~V.} \emph{et~al.}
\newblock \bibinfo{title}{Magnetooptics of exciton rydberg states in a
  monolayer semiconductor}.
\newblock \emph{\bibinfo{journal}{Phys. Rev. Lett.}}
  \textbf{\bibinfo{volume}{120}}, \bibinfo{pages}{057405}
  (\bibinfo{year}{2018}).

\bibitem{Combescot2017}
\bibinfo{author}{Combescot, M.}, \bibinfo{author}{Combescot, R.} \&
  \bibinfo{author}{Dubin, F.}
\newblock \bibinfo{title}{Bose--einstein condensation and indirect excitons: a
  review}.
\newblock \emph{\bibinfo{journal}{Rep. Prog. Phys.}}
  \textbf{\bibinfo{volume}{80}}, \bibinfo{pages}{066501}
  (\bibinfo{year}{2017}).

\end{thebibliography}


\begin{thebibliography}{1}
\expandafter\ifx\csname url\endcsname\relax
  \def\url#1{\texttt{#1}}\fi
\expandafter\ifx\csname urlprefix\endcsname\relax\def\urlprefix{URL }\fi
\providecommand{\bibinfo}[2]{#2}
\providecommand{\eprint}[2][]{\url{#2}}

\bibitem{Wang2017}
\bibinfo{author}{Wang, Z.}, \bibinfo{author}{Zhao, L.}, \bibinfo{author}{Mak,
  K.~F.} \& \bibinfo{author}{Shan, J.}
\newblock \bibinfo{title}{Probing the spin-polarized electronic band structure
  in monolayer transition metal dichalcogenides by optical spectroscopy}.
\newblock \emph{\bibinfo{journal}{Nano Lett.}} \textbf{\bibinfo{volume}{17}},
  \bibinfo{pages}{740--746} (\bibinfo{year}{2017}).

\bibitem{Mak2013}
\bibinfo{author}{Mak, K.~F.} \emph{et~al.}
\newblock \bibinfo{title}{Tightly bound trions in monolayer {MoS}$_2$}.
\newblock \emph{\bibinfo{journal}{Nat. Mater.}} \textbf{\bibinfo{volume}{12}},
  \bibinfo{pages}{207--211} (\bibinfo{year}{2013}).

\bibitem{Ganchev2015}
\bibinfo{author}{Ganchev, B.}, \bibinfo{author}{Drummond, N.},
  \bibinfo{author}{Aleiner, I.} \& \bibinfo{author}{Falko, V.}
\newblock \bibinfo{title}{Three-particle complexes in two-dimensional
  semiconductors}.
\newblock \emph{\bibinfo{journal}{Physical review letters}}
  \textbf{\bibinfo{volume}{114}}, \bibinfo{pages}{107401}
  (\bibinfo{year}{2015}).

\bibitem{Jadczak2017}
\bibinfo{author}{Jadczak, J.} \emph{et~al.}
\newblock \bibinfo{title}{Probing of free and localized excitons and trions in
  atomically thin {WS}e$_2$, {WS}$_2$, {M}o{S}e$_2$ and {M}o{S}$_2$ in
  photoluminescence and reflectivity experiments}.
\newblock \emph{\bibinfo{journal}{Nanotechnology}}
  \textbf{\bibinfo{volume}{28}}, \bibinfo{pages}{395702}
  (\bibinfo{year}{2017}).

\bibitem{Latini2017}
\bibinfo{author}{Latini, S.}, \bibinfo{author}{Winther, K.~T.},
  \bibinfo{author}{Olsen, T.} \& \bibinfo{author}{Thygesen, K.~S.}
\newblock \bibinfo{title}{Interlayer excitons and band alignment in
  {MoS}$_2$/h{BN}/{WSe}$_2$ van der {W}aals heterostructures}.
\newblock \emph{\bibinfo{journal}{Nano Lett.}} \textbf{\bibinfo{volume}{17}},
  \bibinfo{pages}{938--945} (\bibinfo{year}{2017}).

\end{thebibliography}
\clearpage

\section*{Acknowledgments}
This work was supported by the EC-FET European Graphene Flagship (no.785219), the ATOMOPTO project (TEAM programme of the Foundation for Polish Science co-financed by the EU within the ERDFund), the European Research Council (Synergy Grant Hetero2D), the Royal Society (UK), the Royal Academy of Engineering (UK), the Engineering and Physical Sciences Research Council (UK) grants EP/N010345/1, EP/S019367/1, EP/P026850/1, the US Army Research Office (W911NF-16-1-0279), the Elemental Strategy Initiative conducted by the MEXT, Japan and the CREST (JPMJCR15F3), JST.

\section*{Author Contributions}
J.B. carried out the optoelectronic measurements, analyzed and interpreted experimental data, participated in discussions, developed the model and wrote the draft of the manuscript. A.K. produced experimental devices, participated in the measurements, analyzed and interpreted experimental data, contributed to writing the manuscript, J.H. produced experimental devices and participated in the measurements, F.W. produced experimental devices, M. R. M. participated in the measurements, K.W. and T.T. grew high-quality hBN, C.F., A.W., M.D., V.I.F. and A.G. participated in discussions, K.S.N. and M.P. initiated the project, analyzed experimental data, participated in discussions, contributed to writing the manuscript.

\section*{Competing Interests}
The authors declare no competing interests.

\end{document}

\thispagestyle{fancy}

%\preprint{}


\title{ {\Large Supplementary Information} \\ ~ \\~  Upconverted electroluminescence via Auger scattering of interlayer excitons in van der Waals heterostructures}


\author{Johannes Binder et al.}

\maketitle

\subsection*{Supplementary Note 1: Studied devices, their I-V characteristics and comments on their performance}

\begin{figure}[H]
    \centering
        \includegraphics[width=0.85\textwidth]{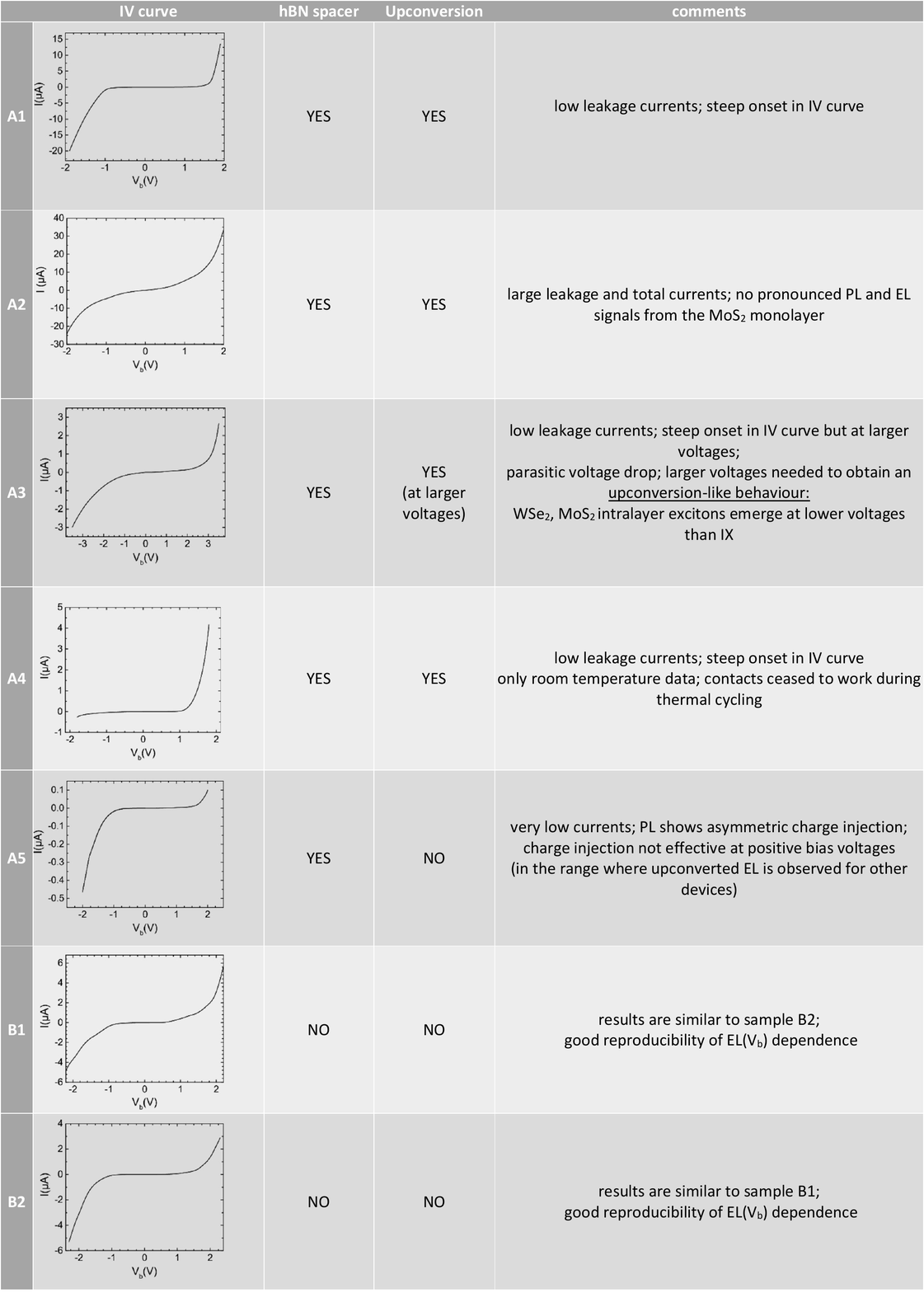}
    \caption{Current-voltage characteristics and additional information for all seven devices studied.}
    \label{fig:IV_A1}
\end{figure}

Altogether seven devices were studied (see
Supplementary Figure~\ref{fig:IV_A1}). Five samples (A1, A2, .. A5)
comprise an hBN spacer between the TMD monolayers, whereas the two other
samples (B1 and B2) were fabricated without an hBN spacer. Each device was electrically tested and current-voltage (I-V)
characteristics (see Supplementary Figure~\ref{fig:IV_A1}) were recorded. The I-V curves for
all seven samples display tunneling diode operation, however the actual
quality of the device performance, mostly determined by leakage currents, varies from sample to sample. Close to optimal is the device A1,
with low leakage currents. Device A4 has only been tested at
room temperature due to contact failure while cooling down the sample. The I-V
characteristic of device A5 is quite distinct from the characteristics of the
other samples. It is asymmetric and displays surprisingly low currents. Device A5 is the only sample with an hBN
spacer that did not show upconverted emission. Both samples without an hBN spacer (B1 and B2) did also not show any upconverted
emission. These two samples display very similar results when studied with electroluminescence (EL), but also photoluminescence (PL) and reflectance contrast (RC) experiments.

\clearpage
\subsection*{Supplementary Note 2: Upconverted emission in other devices}
In this section we present the EL as a function of bias voltage
for other samples that show upconverted emission. As shown in Supplementary Figure~\ref{fig:MQW}, sample A2 is another device
which clearly shows the sub-bandgap EL. We note, however, that in
this sample the emission, both in EL and PL, is only due
to the WSe$_2$ monolayer whereas no signal associated with MoS$_2$ monolayer could
be observed.

\begin{figure}[H]
    \centering
        \includegraphics[width=0.8\textwidth]{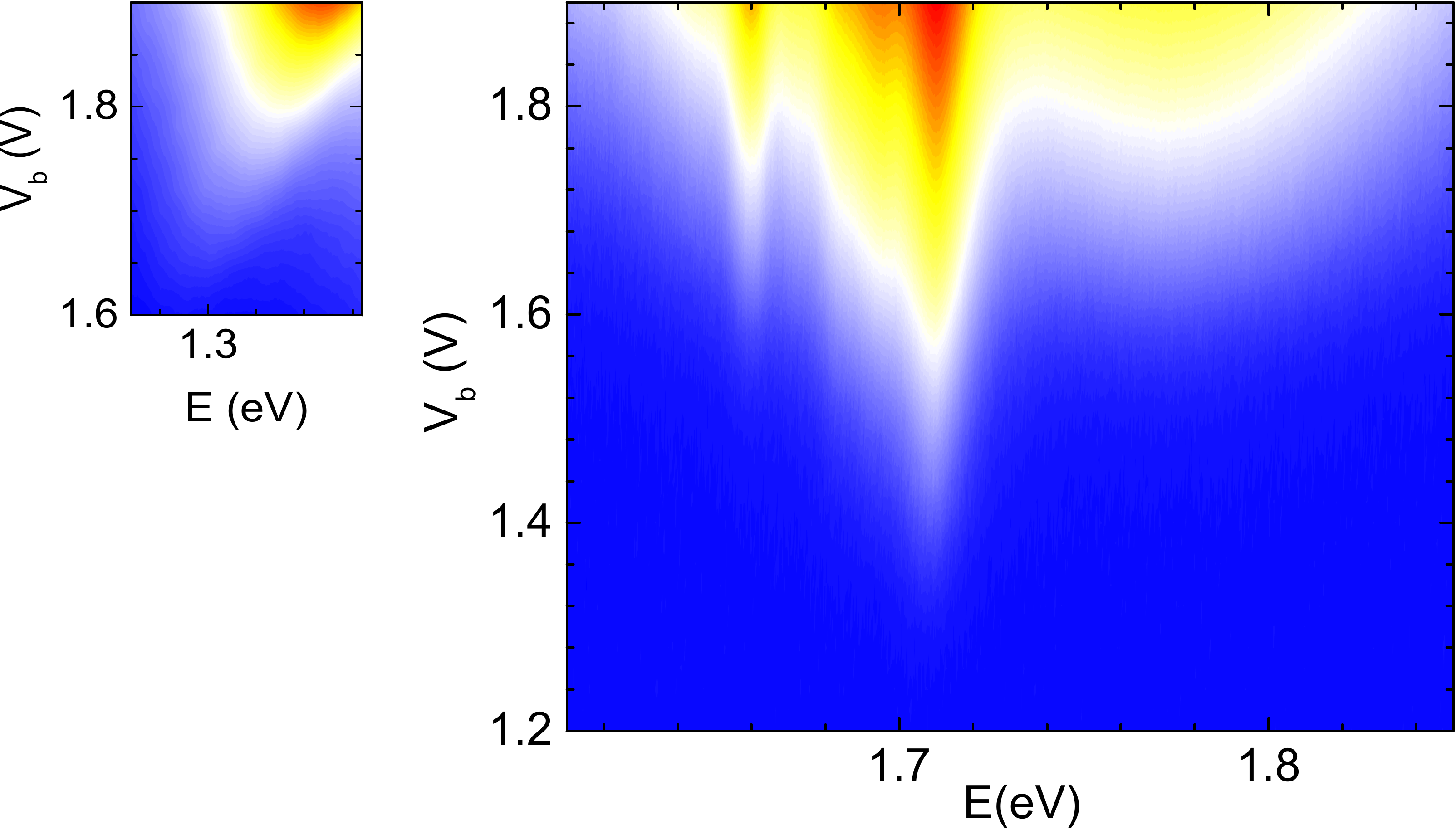}
    \caption{Sample A2 with a monolayer hBN spacer showing the upconverted EL. For this device no intense signal from MoS$_2$ has been observed, both in EL and PL.
    The monolayer-like emission, including the upconverted emission, is mainly associated to the WSe$_2$ monolayer.
    In agreement with the results shown in the main text, the emission of the IX is emerging at larger voltages $\sim 1.6$~V as compared to the WSe$_2$ signal at $\sim 1.3$~V.}
    \label{fig:MQW}
\end{figure}

To clearly demonstrate the onset of the "sub-bandgap" EL, we show a
series of EL spectra measured in the vicinity of the threshold
voltage, both for sample A1 (from the main text) and sample A2. For
device A1 we start to see EL corresponding to the A-exciton resonance of the WSe$_2$ and MoS$_2$ monolayers at voltages as
low as V$_{\text{b}}=1.30$~V. For device A2, even at voltages of
V$_{\text{b}}=1.28$~V electroluminescence of the WSe$_2$ monolayer can be observed.

\begin{figure}[H]
    \centering
        \includegraphics[width=0.9\textwidth]{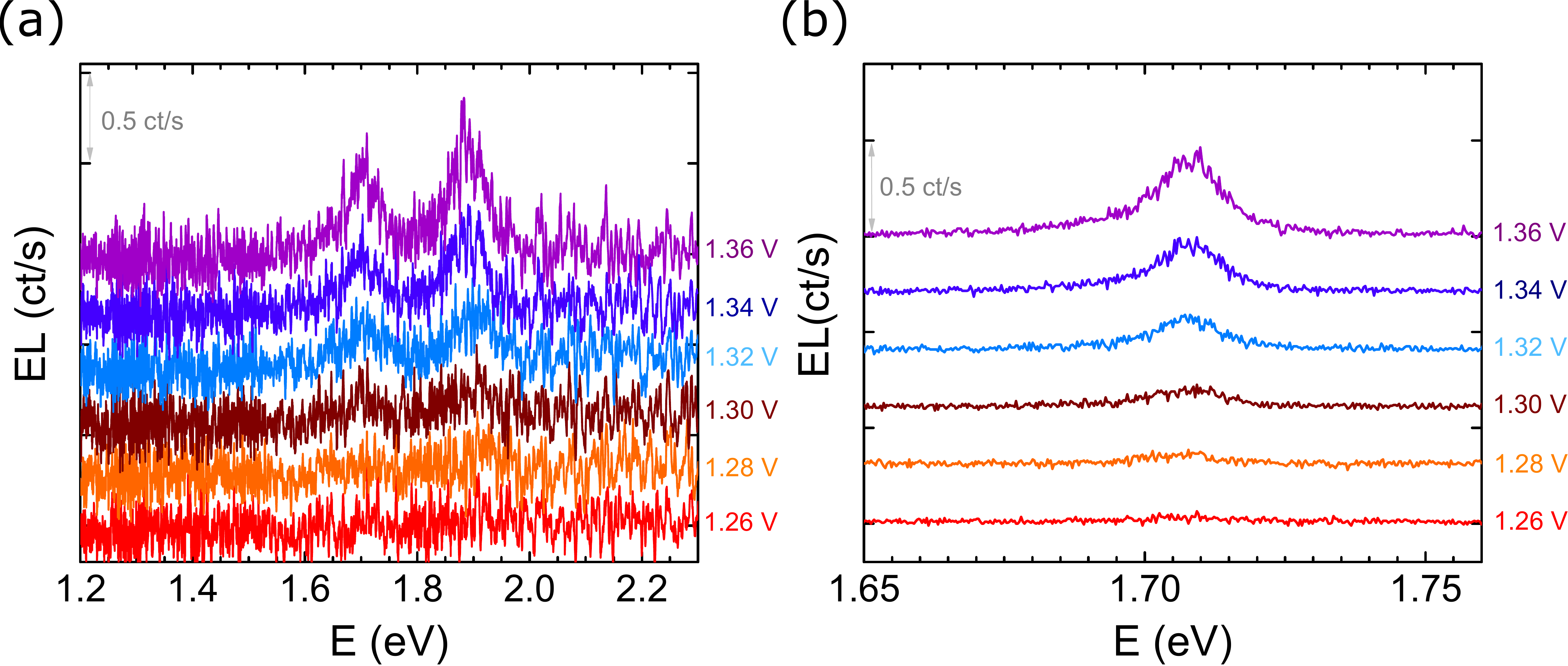}
    \caption{Comparison of the EL spectra close to the onset of the upconverted emission for (a) sample A1 and (b) sample  A2. For sample A1 (A2) EL can be observed at voltages as low as  V$_{\text{b}}=1.30$~V (V$_{\text{b}}=1.28$~V). The spectra are shifted vertically for clarity.}
    \label{fig:threshold}
\end{figure}

\clearpage
The EL of device A3, shown in Supplementary Figure \ref{fig:db}, shows similar
characteristic features like device A1, discussed in the main
text. The EL of the WSe$_2$ and MoS$_2$ contributions emerge at lower bias voltages
than the IX emission and the onset of the emission due to the A-exciton resonance of WSe$_2$ and MoS$_2$ is
virtually the same. The only major difference is that larger
voltages are needed to observe emission. We believe that the
larger voltages can be explained by an additional voltage drop on
our device, so that the overall applied voltage does not
correspond to the voltage dropping across the active area. This
conclusion agrees well with the currents at larger voltages (V$_{\text{b}}
>$ 3~V) observed in the IV characteristics, which are similar to
the currents for other devices biased at lower voltages (V$_{\text{b}} <$
2~V), see Supplementary Figure \ref{fig:IV_A1}. Given the fact that besides the
parasitic voltage drop all other characteristics are similar, we
refer to this behavior as "upconversion-like". Please note that
for the devices that did not show upconverted emission, (i.e., the
device B1 from the main text and sample B2 discussed in the next
section) the EL of the monolayer TMDs emerges according to the magnitude of
the bandgap, but the IX appears first at lower bias voltages.
%Thus
%the behavior of device A3 is very distinct from the behavior se
%devices.

\begin{figure}[H]
    \centering
        \includegraphics[width=0.8\textwidth]{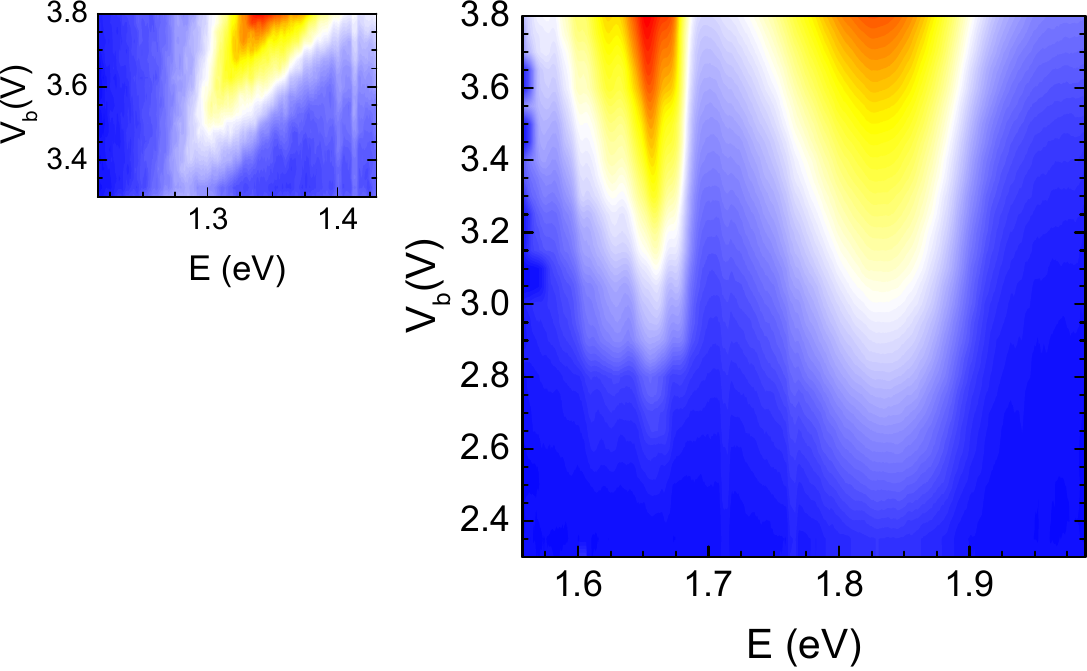}
    \caption{Sample A3 with a monolayer hBN spacer. Similar to the other samples for which upconversion is observed the signals from WSe$_2$ and MoS$_2$ emerge at lower voltages
    than the emission corresponding to the IX.
    Most probably the voltage does not solely drop across the active area. This explains the larger voltages needed to observe the effects.}
    \label{fig:db}
\end{figure}

\begin{figure}
    \centering
        \includegraphics[width=0.5\textwidth]{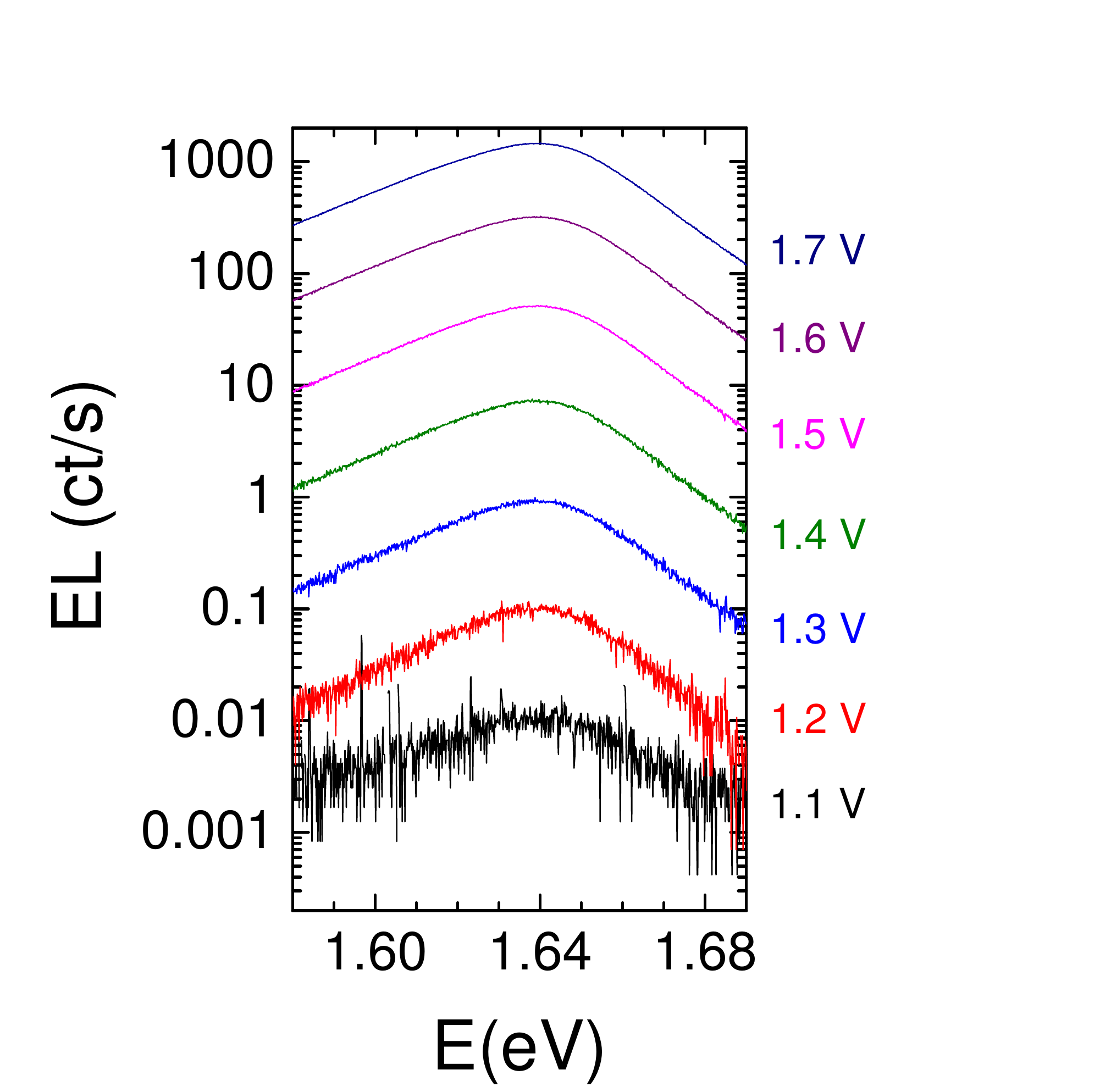}
    \caption{Sample A4 with a monolayer hBN spacer showing upconversion.
    For this sample only measurements at room temperature were possible.
    At room temperature a broad upconverted peak corresponding to WSe$_2$ can be clearly seen.
    EL at energies around 1.64~eV is observed at bias voltages as low as 1.1~V.}
    \label{fig:D1}
\end{figure}

Supplementary Figure \ref{fig:D1} presents results for device A4. Unfortunately,
this device could only be measured at room temperature, since the
electric contacts ceased to operate during thermal cycling. The
initial measurements at room temperature, however, already show a
broad upconverted emission corresponding to the A-exciton resonance of WSe$_2$ at around 1.64~eV for
voltages as low as 1.1~V. The I-V characteristics of this device
show a steep onset and low leakage currents as shown in Supplementary Figure
\ref{fig:IV_A1}.

\clearpage

\subsection*{Supplementary Note 3: Electroluminescence of devices not showing upconverted emission}

Device B2, with a similar design to sample B1
discussed in the main text (no hBN spacer), shows practically the same response as B1 in
our experiments. The "regular" not upconverted behavior of the EL
signal in sample B2 is illustrated in Supplementary Figure~\ref{fig:D2}.
The IX electroluminescence of sample B2 appears at bias voltages
around 1.65~V whereas the monolayer emission due to WSe$_2$ and
MoS$_2$ emerges progressively at higher voltages, ~1.85~V and
2.15~V, correspondingly.

\begin{figure}[H]
    \centering
        \includegraphics[width=0.6\textwidth]{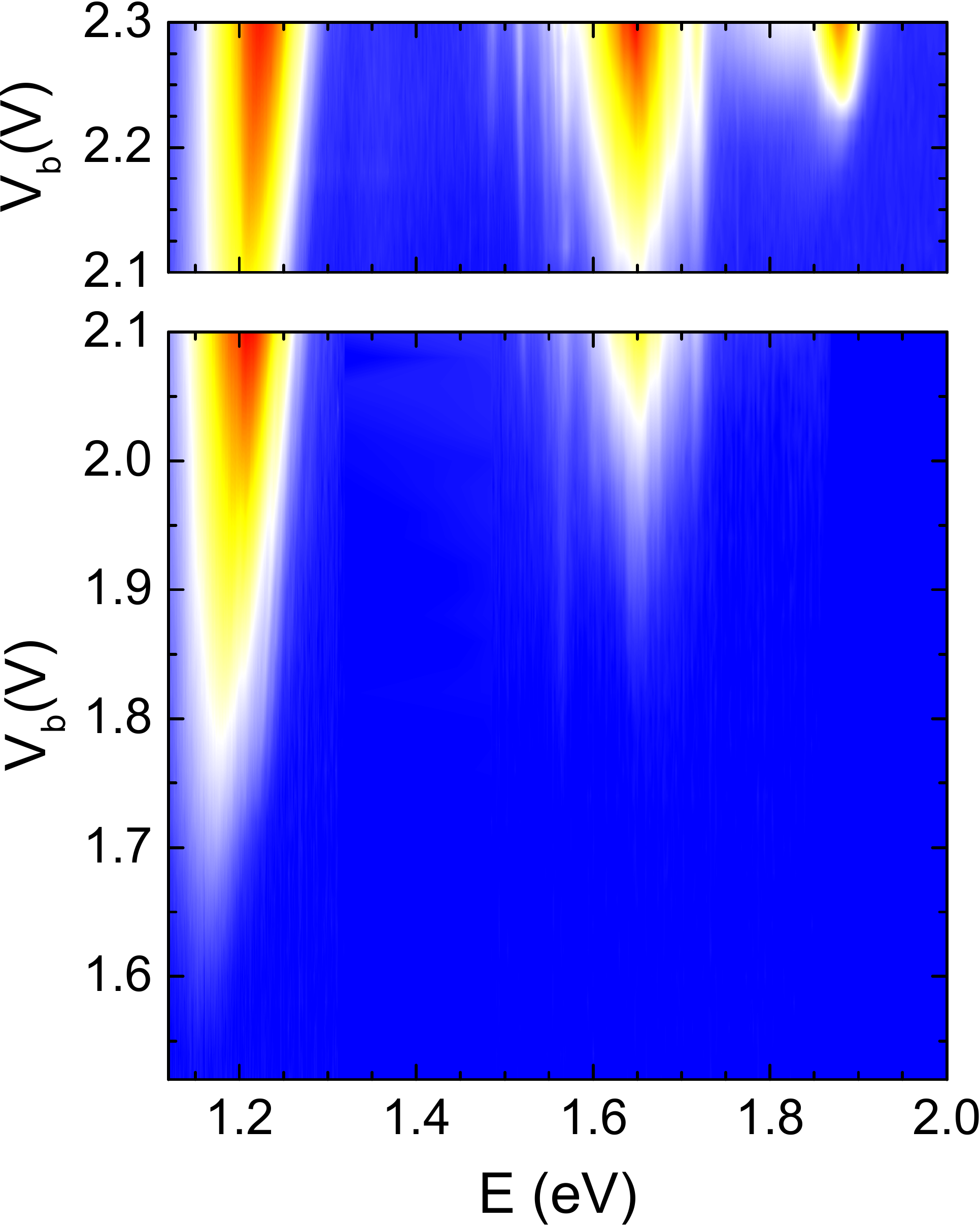}
    \caption{Sample B2 without hBN spacer. Similar to the sample in the main text, no upconverted emission was observed. The evolution of the contributions of the IX, WSe$_2$ and MoS$_2$ is similar to that  for the device B1 discussed in the main text. The two graphs show two consecutive measurements sweeps.}
    \label{fig:D2}
\end{figure}

\begin{figure}[H]
    \centering
        \includegraphics[width=0.7\textwidth]{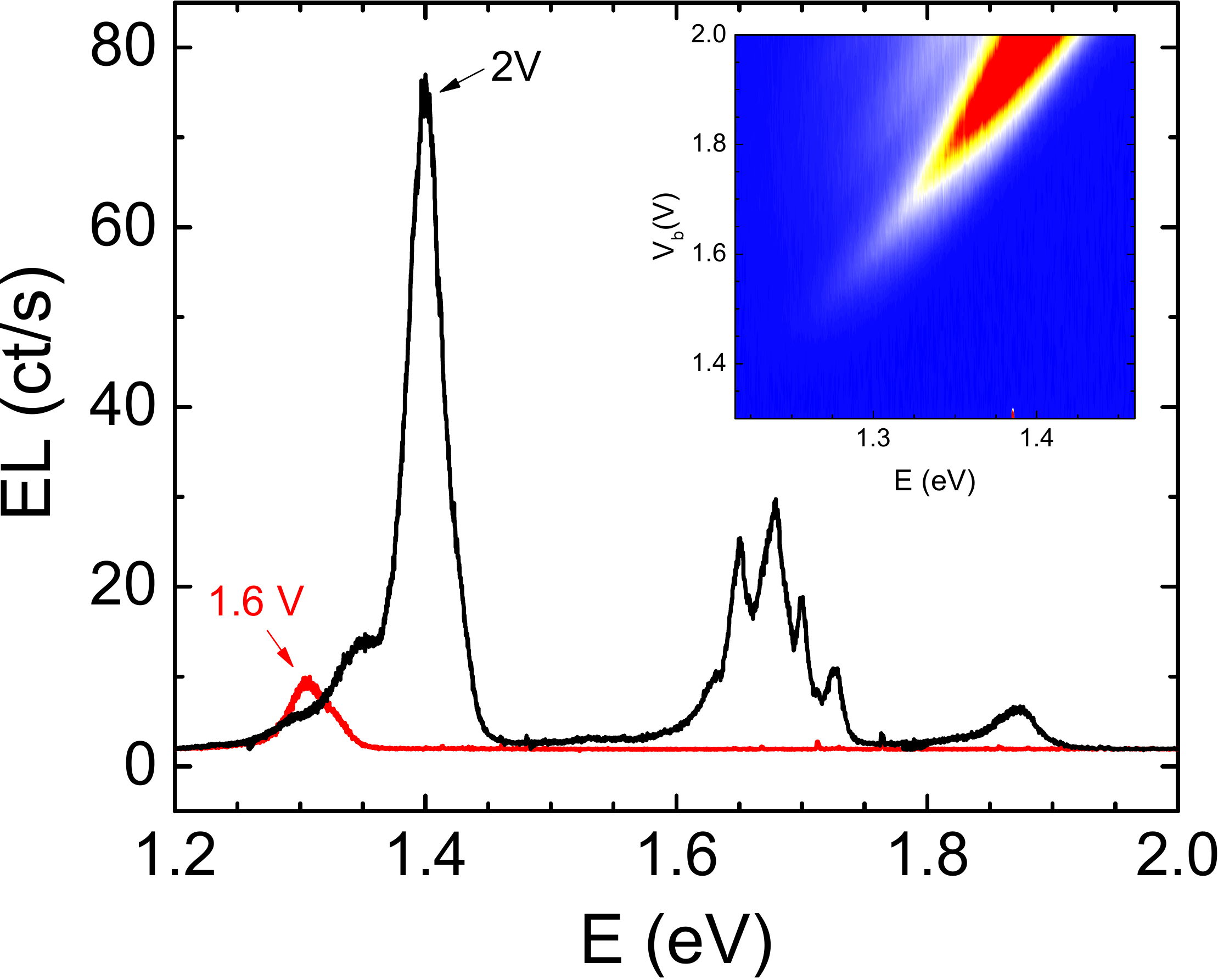}
    \caption{Sample A5 with a monolayer hBN spacer. Two EL spectra for different bias voltages indicate that no upconverted emission was observed for this sample. In contrast to the samples showing upconverted emission, the emission from the IX emerge first. This sample showed overall low currents and PL measurements indicated a very asymmetric carrier injection. \textit{Inset:} Map of IX emission as a function of bias voltage.}
    \label{fig:da}
\end{figure}

Supplementary Figure \ref{fig:da} presents the results of EL measurements for device A5. Although this device comprises a monolayer hBN
spacer, no upconverted emission was observed. Device A5 is,
however, very distinct from all other devices studied. It shows
low overall currents and a strong asymmetry of the I-V
characteristics, with larger currents in the negative voltage
direction (Supplementary Figure \ref{fig:IV_A1}). PL measurements as a function
of bias voltage indicate a very asymmetric charge injection, with
a trion peak emerging only in one voltage direction. These
findings show that the injection through one graphene layer is
much less effective and it can hence be understood that such a
large difference between electron and hole concentrations does not
allow to establish IX densities large enough to observe the
excitonic Auger process and hence upconverted emission.

\begin{figure}[H]
    \centering
        \includegraphics[width=0.95\textwidth]{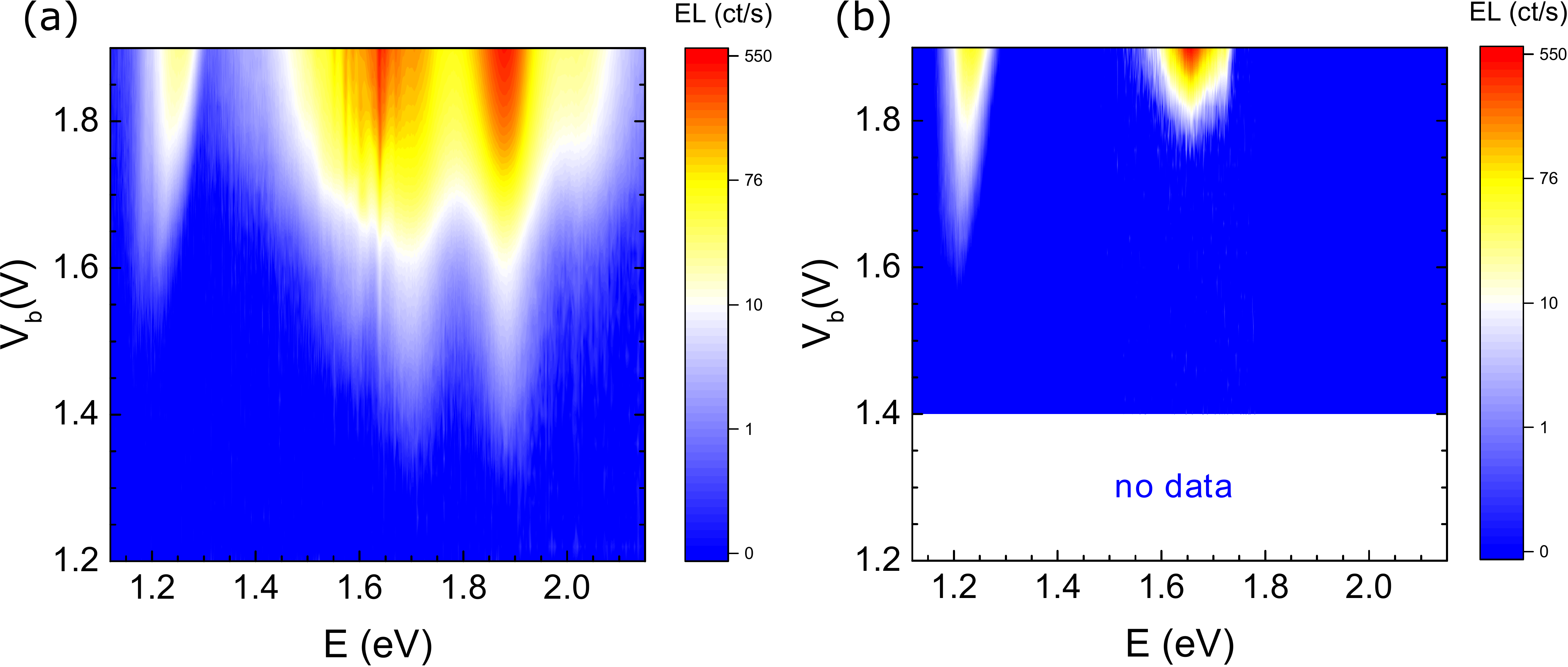}
    \caption{Side by side comparison of the contour maps of the EL spectra measured for devices (a) A1 and (b) B1. The Supplementary Figures present the same data as shown in Figures 3 (a) and 2 (a) in the main text, but both with the same contrast (thus restricted range of bias voltages) and in the same spectral range.}
    \label{fig:A1_B1_same_scale}
\end{figure}

Supplementary Figure \ref{fig:A1_B1_same_scale} illustrates the different voltage onsets for intralayer emission for devices A1 and B1 on the same false color and bias voltage scale. The integrated electroluminescence intensities corresponding to these graphs were presented in Figure 3 (c) in the main text.

\clearpage

\subsection*{Supplementary Note 4: Electroluminescence, reflectance contrast and photoluminescence versus injected current/bias voltage: Supplementary data and estimation of interlayer exciton densities}

\subsubsection{EL intensity versus driving current}

\begin{figure}[h]
    \centering
        \includegraphics[width=0.8\textwidth]{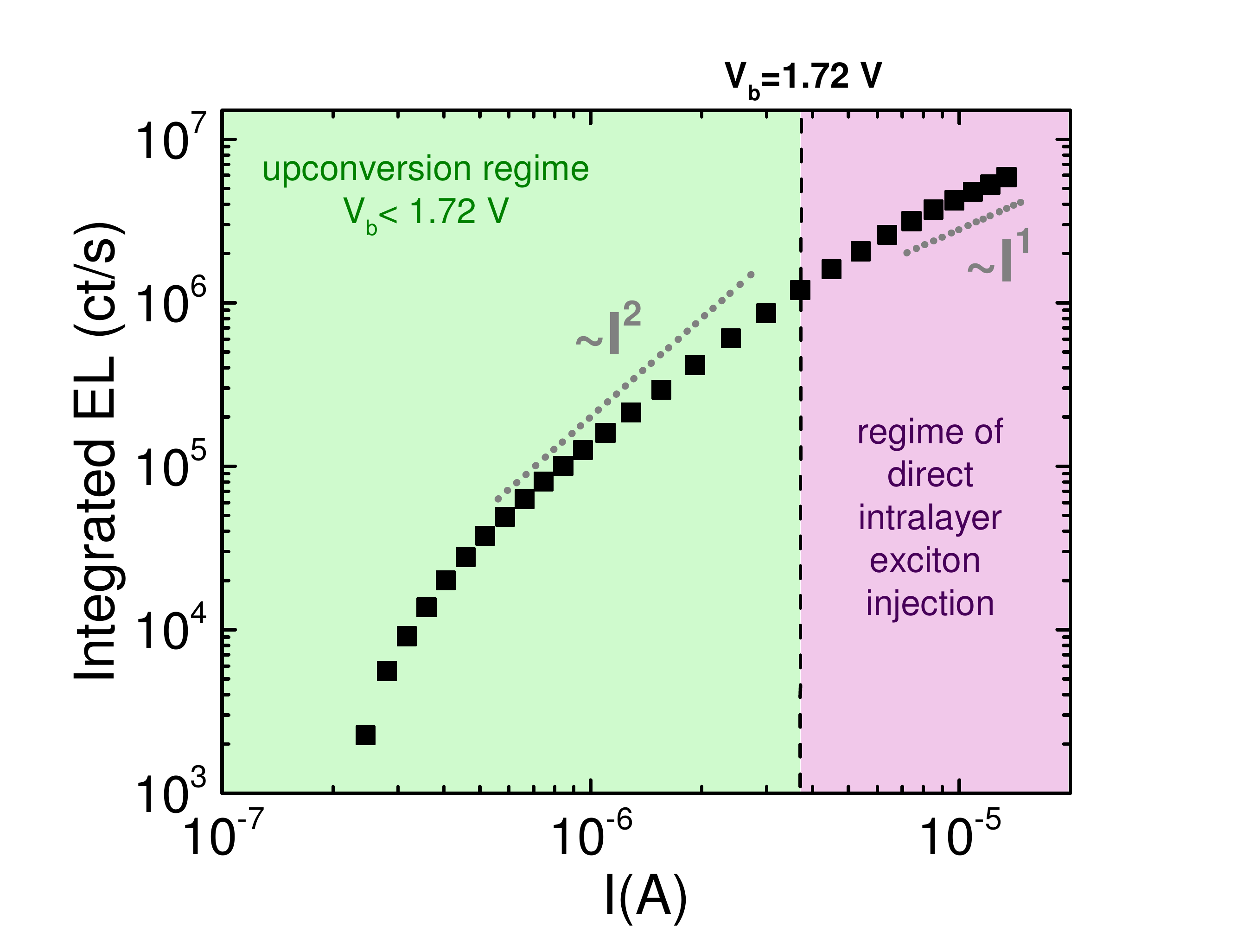}
    \caption{Integrated EL of the intralayer exciton emission as a function of the driving current (log-log plot) for  device A1. Quadratic ($\sim $I$^2$) and linear ($\sim $I$^1$) trends are presented for comparison.}
    \label{fig:Curr_IX_A1_B1}
\end{figure}

Supplementary Figure \ref{fig:Curr_IX_A1_B1} presents the integrated intralayer
A-exciton EL of both, MoS$_2$ and WSe$_2$, as a function of
driving current for sample A1 in a log-log representation. The
Figure depicts two regimes: the first regime, below a voltage of
V$_{\text{b}}$=1.72~V (corresponding to the intralayer exciton emission
energy of 1.72~eV), corresponds to upconverted EL. Above this
voltage, direct injection into intralayer excitonic states becomes
possible and we enter the second regime. Supplementary Figure
\ref{fig:Curr_IX_A1_B1} shows a super-linear trend for the
upconverted EL. Super-linear trends are expected due to the many
body nature of the process. At larger voltages, in the second
regime, we observe an almost linear behavior, which agrees well
with the expectations of direct charge carrier injection of
intralayer excitons.

Although the extracted quantities are reasonable and within the
expectations they have to be taken with care. The problem is that
the driving current (x-axis of the graph) is strongly influenced
by leakage currents which especially gain importance for low
injection. In Supplementary Figure \ref{fig:Curr_IX_A1_B1} one can observe this
behavior at low currents (I$~< 0.5$~ $\mu$A). In this regime one
obtains larger exponents, which constitute  an overestimation of
the actual super-linear trend. The current-voltage characteristics
for sample A1 (Supplementary Figure \ref{fig:IV_A1}) show a very pronounced
diode-like behavior with small currents at low voltages and a very
steep onset at the threshold for direct carrier injection.
However, not all the devices measured do show such an almost ideal
behavior (see Supplementary Figure \ref{fig:IV_A1}). The observed leakage leads
to the above described overestimation of the super-linearity also
for large voltages and currents. For device B1 for example, we can
extract a dependency of $\sim $I$^{2.8}$ in the regime of direct
carrier injection, which is in contrast to the almost linear
behavior seen in this regime for sample A1, only enabled by the
excellent IV characteristics of this sample. In our two-terminal
devices it is very difficult to establish the actual nature of the
leakage contributions, given the many possible different leakage
channels for which in general one cannot assume a linear
dependence on voltage. Although we observe a super-linear trend
for the upconverted intralayer EL and an almost linear behavior in
the direct injection regime for sample A1, an estimation relying
on this measure may, however, be questionable. Optionally, we
propose two other approaches to estimate the carrier concentration
based on additional optoelectronic measurements.

\subsubsection{Carrier injection probed with RC measurements}
\label{PLtrion} 
\begin{figure}
    \centering
        \includegraphics[width=0.7\textwidth]{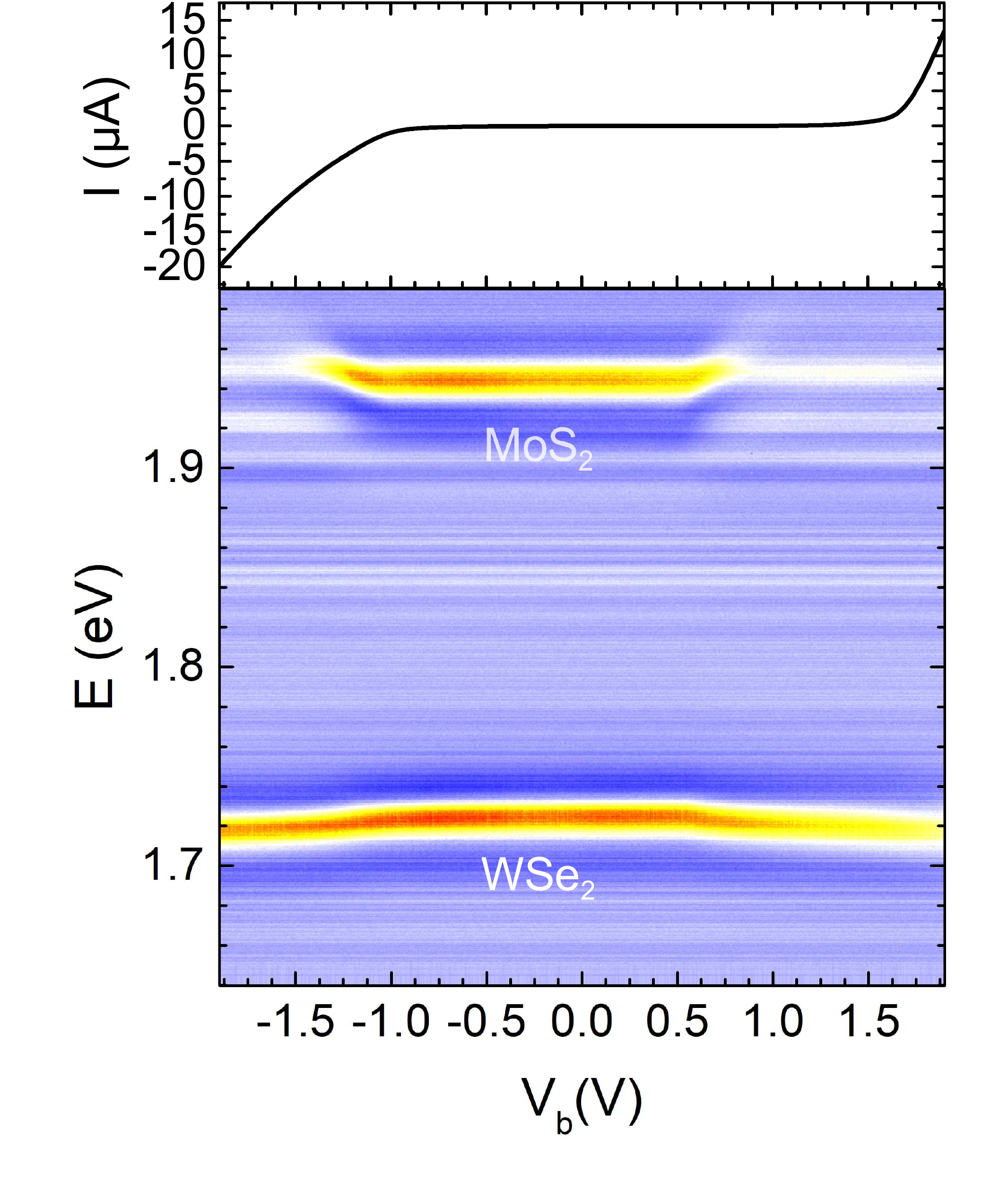}
    \caption{Reflectance contrast measurements as a function of bias voltage for device A1.
    The false color map shows the 2nd derivative of the measured RC signal. The A exciton resonance could be resolved for both MoS$_2$ and WSe$_2$.
    A clear change in the spectra is visible at about 0.6 - 0.7~V for both monolayers. The upper panel presents the I-V characteristics recorded simultaneously with the RC data.}
    \label{fig:d5_merged_RCs}
\end{figure}

Supplementary Figure \ref{fig:d5_merged_RCs} presents the
results of reflectance contrast (RC) measurements as a function of
the applied bias voltage together with the simultaneously measured
IV curve for device A1. The RC spectra were obtained by focusing a
white light spot on the active area of the device followed by the
subtraction of the reference spectrum of the bare SiO$_2$/Si
substrate. In order to illustrate the changes more markedly we
plot the 2nd derivative of the RC signal.

A clear change in the absorption spectrum
can be observed at about 0.6 - 0.7~V. The A-exciton resonance of
monolayer MoS$_2$ is quenched in this region and a kink can be
observed for the A-exciton resonance of monolayer WSe$_2$. Due to
the band alignment we know that electrons are injected via
tunneling into the MoS$_2$ monolayer and holes into the WSe$_2$
monolayer. The quenching of the A-exciton resonance of MoS$_2$
shows the common behavior, for example also observed in dual-gated
MoSe$_2$ monolayers \cite{Wang2017}. The quenching of the A-exciton in
the WSe$_2$ monolayer is somewhat less abrupt but the progressive
change of its spectrum is clearly indicative of the effective
injection of holes into the valence band of WSe$_2$.
The upper panel in Supplementary Figure \ref{fig:d5_merged_RCs} shows the
current-voltage characteristics measured simultaneously with the
the RC measurement. For negative voltages, we see a good
correlation between the current onset and changes in the RC.
However, in the positive direction, for which upconverted emission
was observed, the current remains very small although the optical
absorption measurement shows a clear change in carrier
concentration. This behavior can be taken as a sign of carrier
accumulation into states with a long lifetime resulting in a small
current caused by carrier recombination. At voltages above 1.7~V,
a direct carrier injection into intralayer excitonic states
becomes possible and hence we expect a large recombination current
component due to the effective radiative recombination and intense
light emission in this region. The latter behavior is similar to
conventional light emitting diodes.

\subsubsection{PL versus bias voltage and estimation of densities of the injected carriers}

\begin{figure}
    \centering
        \includegraphics[width=0.7\textwidth]{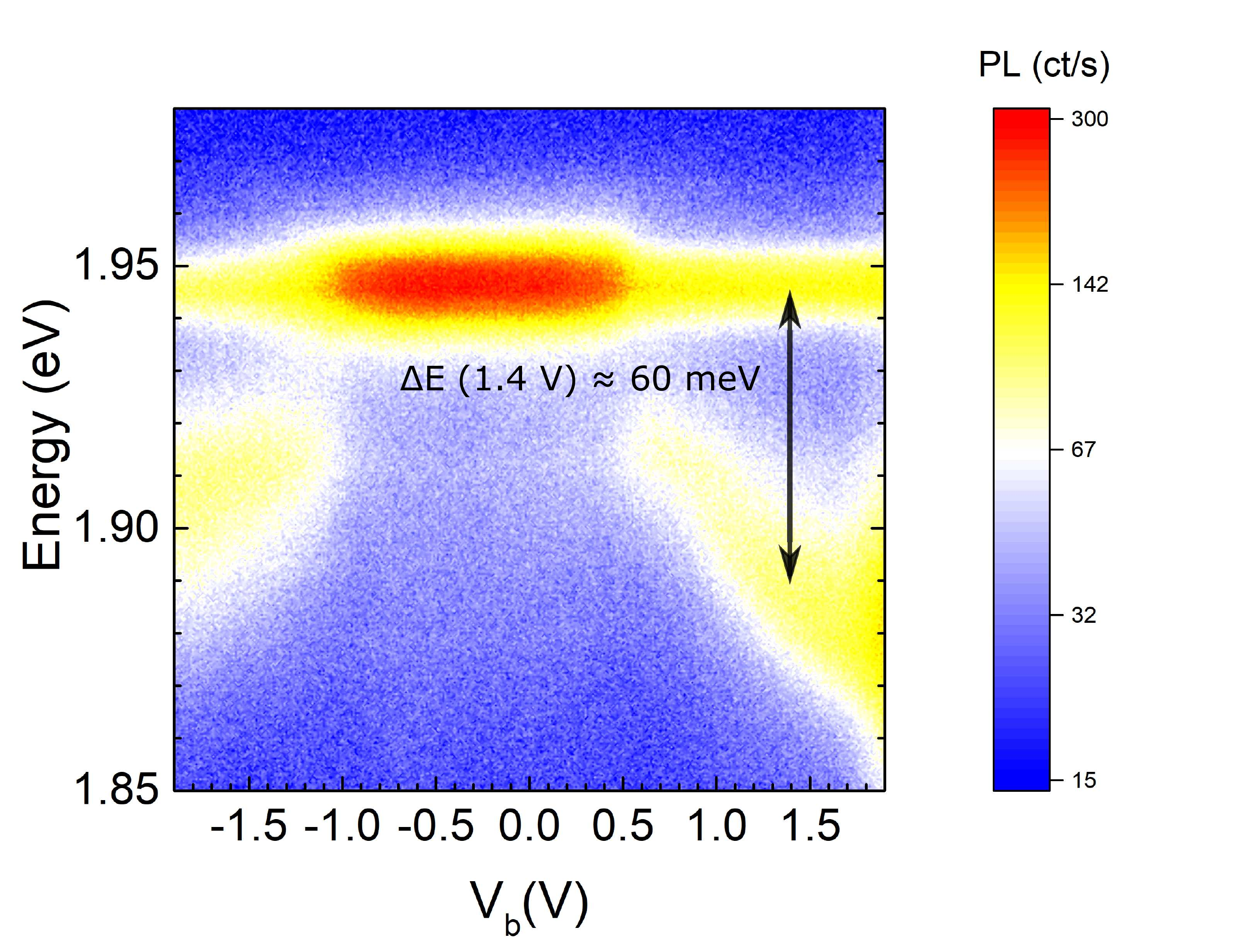}
    \caption{Photoluminescence as a function of bias voltage for device A1 for an excitation wavelength of 514~nm. The false colour map depicts the behavior of the A-exciton resonance of MoS$_2$. At a voltage of about 0.6 - 0.7~V a trion peak emerges and the free excitonic emission is quenched.}
    \label{fig:d5_PL_contour_log}
\end{figure}

Besides the RC measurements, which provide us with information
about light absorption, we also performed the photoluminescence
(PL) measurements as a function of bias voltage. Supplementary Figure
\ref{fig:d5_PL_contour_log} presents such a measurement for device
A1.

\begin{figure}
    \centering
        \includegraphics[width=0.70\textwidth]{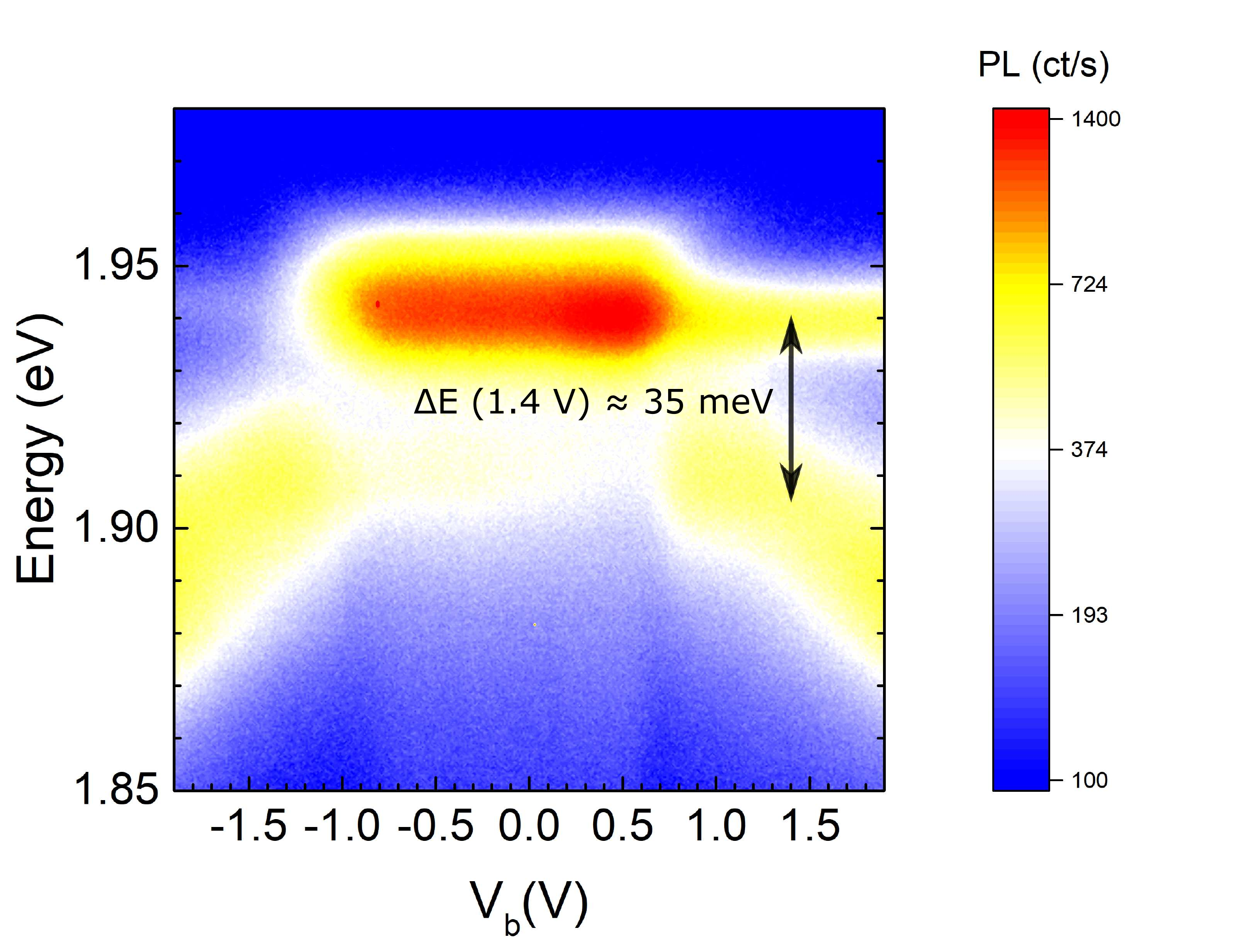}
    \caption{Photoluminescence as a function of bias voltage for device B1 for an excitation wavelength of 514~nm. The false colour map depicts the behavior of the A-exciton resonance of MoS$_2$. At a voltage of about 0.6 - 0.7~V a trion peak emerges and the free excitonic emission is quenched.}
    \label{fig:d4_PL_cont_log_Xinv}
\end{figure}

Supplementary Figures \ref{fig:d5_merged_RCs} and
\ref{fig:d5_PL_contour_log} are in very good agreement. They show
a quenching of the free exciton contribution at a voltage of about
0.6 - 0.7~V. This quenching allows us to conclude that at this voltage
charge carriers are started to be effectively injected into states
with a long lifetime yielding to a large charge build-up across
the interlayer bandgap. Moreover, the PL of the MoS$_2$ monolayer
reveals a trion resonance at voltages exceeding 0.6 - 0.7~V, which
further indicates that charge carriers are injected. Supplementary Figure
\ref{fig:d4_PL_cont_log_Xinv} presents a PL sweep for device B1,
without hBN spacer, hence for a sample that does not show
upconversion. Again a trion peak emerges, at voltages of about 0.6 - 0.7~V.

Both RC and PL measurements provide evidence for charge carrier
injection at voltages well below the thresholds for intralayer
exciton injection ($\sim 1.7$~V for WSe$_2$) in agreement to the
large expected charge build-up described in the main text. To
obtain an estimation of the carrier concentration we use the
dependencies presented in Ref.~\onlinecite{Mak2013} allowing to
translate the shift in energy of the trion $\Delta $E$=$E$_{\text{A}}-$E$_{\text{A'}}$ ,
where E$_{\text{A}}$ is the energy of the neutral exciton and E$_{\text{A'}}$ the
energy of the trion, into a shift of the Fermi level E$_{\text{f}}$.

\begin{equation}
    \Delta E (V)= 1.2 \cdot E_f(V) + E_b
\end{equation}

where E$_{\text{b}}$ is the trion binding energy. Assuming an effective
mass for electrons of m$^{*}=0.35~$m$_{\text{e}}$ \cite{Mak2013} we can
calculate the carrier concentration from the Fermi level with the
formula:

\begin{equation}
    n(V)= \frac{\Delta E (V) -E_b}{1.2} \cdot \frac{m^{*}}{\pi \hbar^2}\approx \frac{\Delta E (V) -E_b}{1.2} \cdot 1.46 \cdot 10^{14} \frac{1}{eV \cdot cm^{-2}.}
\end{equation}

From Supplementary Figures \ref{fig:d5_PL_contour_log} and \ref{fig:d4_PL_cont_log_Xinv} we can extract $\Delta $E$_{\text{A1}} \sim 60$~meV and $\Delta $E$_{\text{B1}} \sim 35$~meV at a voltage of 1.4~V and a binding energy of roughly 30~meV in agreement with literature data\cite{Ganchev2015,Jadczak2017}, which  yields a carrier concentration in MoS$_2$ of n$_{\text{A1}}\sim 3.7 \cdot 10^{12}~$cm$^{-2}$ at a voltage of 1.4~V. A lower concentration of n$_{\text{B1}}\sim 6.1 \cdot 10^{11}~$cm$^{-2}$ can be estimated for sample B1, without hBN spacer, not showing upconversion.

\subsubsection{Blueshift of the interlayer exciton with applied voltage and estimation of interlayer exciton density}
\label{IXshift} The second measure allowing to estimate the
carrier concentration is the blue-shift of the IX, which is a
result of the relative band movements caused by an increasing
electric field. The PL and RC measurements indicate that at
voltages as low as 0.6 - 0.7~V charge carriers are started to be
injected, hence we can conclude that at a voltage of about 1.5~V,
for which the IX emerges in the spectrum, a large charge build-up
must already be present at the interface between the TMDs. Based
on this observation one can assume that the shift of the IX is
mostly governed by the charge carriers at this interface and a
simple parallel plate capacitor model can be used to the estimate
a carrier concentration n:

\begin{equation}
    n=\Delta E_{IX} \cdot \frac{\epsilon_0}{e^2} \cdot \frac{\epsilon_r}{d}
\end{equation}

where $\Delta$ E$_{\text{IX}}$ is the absolute blue-shift, $\epsilon_0$ the vacuum permittivity, $\epsilon_r$ the relative permittivity and d the distance between the capacitor plates. To estimate the carrier concentration we have to determine the absolute blue-shift $\Delta$ E$_{\text{IX}}$. We can extract the energy of the IX as a function of voltage E$_{\text{IX}}$(V) from our EL maps revealing a linear behavior. The only unknown parameter is the energy of the IX without injected charge carriers E$_{\text{IX}}$(V$_{\text{0}}$). As mentioned in the main text, no PL of the IX could be observed at zero voltage so we cannot directly measure this quantity. However, the PL and RC measurements provide us with an estimation concerning the minimal voltage threshold for carrier injection into the TMDs. For device A1 this would be around $V_{\text{thres}_{\text{A1}}}=0.6~$V. We can hence use this value and extrapolate the fitted IX positions to this voltage which gives an energy of E$_{\text{IX}_{\text{A1}}}(0.6~V) \approx 1.08~$eV. In the case of device A1 we obtain the simple formula $ \Delta$ E$_{\text{IX}}$(V)$=$E$_{\text{IX}}$(V)$-$E$_{\text{IX}_{\text{A1}}}$(0.6~V). The general formula for all devices is

\begin{equation}
    n(V)=(E_{IX}(V)-E_{IX}(V_{thres})) \cdot \frac{\epsilon_0}{e^2} \cdot \frac{\epsilon_r}{d}
\end{equation}

where E$_{\text{IX}}$(V$_{\text{thres}}$) is the threshold voltage for charge accumulation extracted from PL.

E$_{\text{IX}}$(V) is presented for several devices in the inset of Figure 2 in the main text. The linear trends for devices B1 and B2 have similar slopes, however the trend for device B1 is shifted to larger energies. The extrapolation for this device deviates from the others and for an extracted voltage of V$_{\text{thres}_{\text{B1}}}=0.6~$V we obtain E$_{\text{IX}_{\text{B1}}}$(0.6~V)$ \approx 1.11~$eV for the presumptive energy of IX emission without injected carriers. The estimated carrier concentrations for device B2 are very similar, since due to the parallel slope, the difference E$_{\text{IX}}$(V$_{\text{thres}}$) is compensated by lower values of E$_{\text{IX}}$(V). Besides the variation of $\Delta$E$_{\text{IX}}$ the carrier concentration will also be strongly influenced by the ratio of $\epsilon_r$ and d.

\begin{figure}
    \centering
        \includegraphics[width=0.8\textwidth]{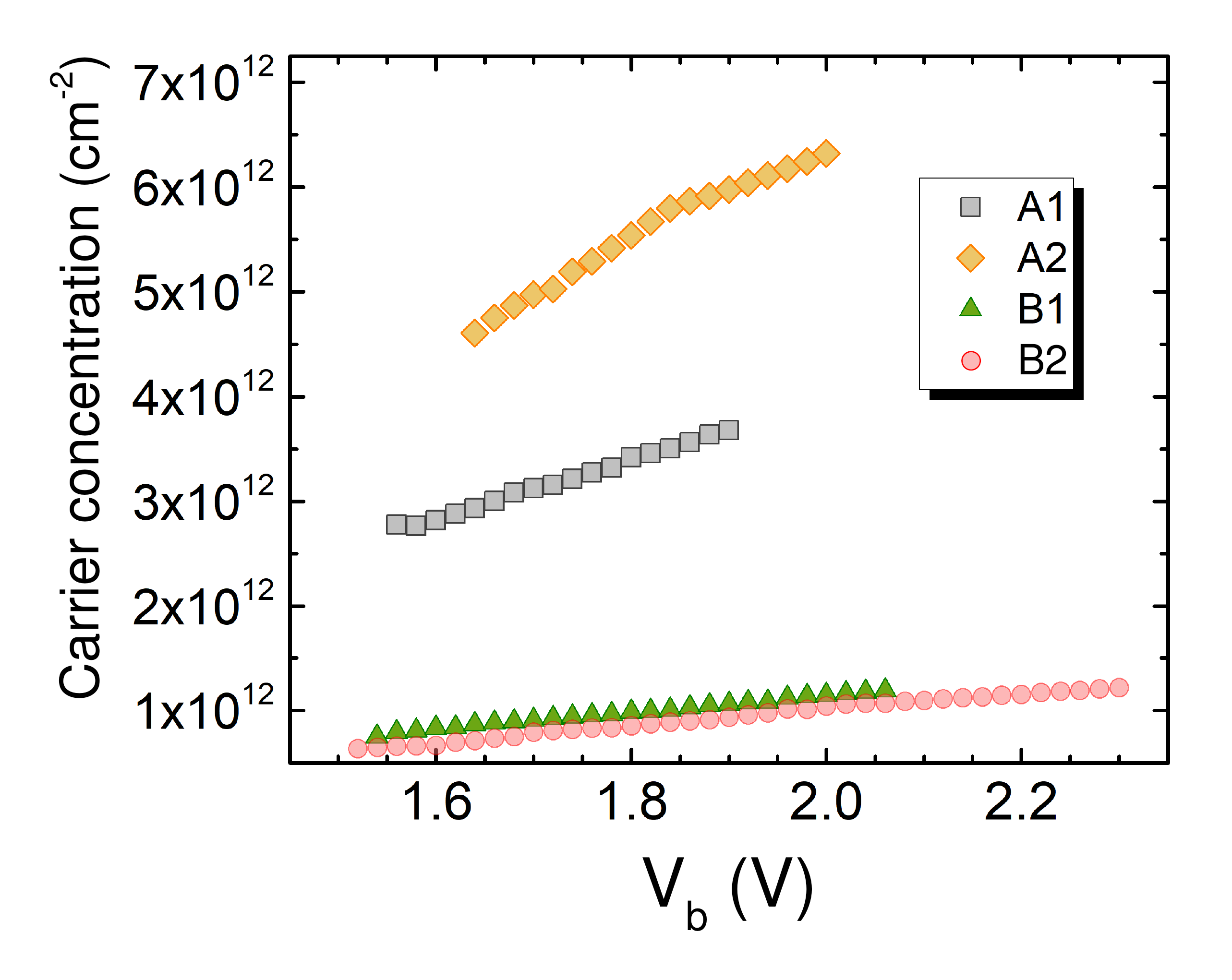}
    \caption{Carrier concentrations as a function of bias voltage. The graph summaries the results of the estimation using the blueshift of the IX emission for devices A1, A2, B1 and B2.}
    \label{fig:Carrier_conc_IX_PL}
\end{figure}

For the distance d we use values from
Ref.~\onlinecite{Latini2017} of d$_{\text{WSe}_{2}/\text{hBN}}=0.528~$nm,
d$_{\text{MoS}_{2}/\text{hBN}}=0.508~$nm and d$_{\text{WSe}_{2}/\text{MoS}_{2}}=0.651~$nm giving a
distance for device group A with an hBN spacer of
d$_{\text{with~hBN}}=0.528~$nm$+0.508~$nm$=1.036~$nm and for device group B
without hBN spacer of d$_{\text{without~hBN}}=0.651~$nm. For samples with
hBN we assume $\epsilon_{r_{\text{with~hBN}}}=4$ and for samples without
hBN we assume a vdW gap, hence $\epsilon_{r_{\text{without~hBN}}}=1$.
This simple estimation yields a carrier concentration of
n$_{\text{A1}}\sim 2.8 \cdot 10^{12}~$cm$^{-2}$ for device A1 and
n$_{\text{B1}}\sim 8.3 \cdot 10^{11}~$cm$^{-2}$ for device B1 at a voltage
of 1.6~V. Supplementary Figure \ref{fig:Carrier_conc_IX_PL} presents the results for devices A1,
A2 and B1, B2 as a function of bias voltage. 

The Figure only
depicts points based on values extracted from the measurement and
no interpolation. It can clearly be seen that both devices with an
hBN spacer show carrier concentrations in the $10^{12}~$cm$^{-2}$
range whereas devices without hBN are in the range of
$10^{11}~$cm$^{-2}$. Both presented methods, i.e. the MoS$_2$ trion
shift and IX blueshift, are in good agreement and show that the
charge carrier concentrations are much larger in the case of the
samples with an hBN spacer.

\clearpage

\subsection*{Supplementary References}

\bibliographystyle{naturemag} 
\bibliography{references}
\clearpage